\documentclass[letterpaper,twocolumn,10pt]{article}
\usepackage{usenix-2020-09}

\usepackage{tikz}
\usepackage{amsmath}
\usepackage{amssymb}
\usepackage{algorithm}
\usepackage[noend]{algpseudocode}
\usepackage{booktabs}
\usepackage{subcaption}
\usepackage{xspace}
\usepackage{enumitem}

\DeclareMathOperator*{\argmin}{arg\,min}
\newcommand{\systemname}{\textsc{Seer}\xspace}

\begin{document}
%-------------------------------------------------------------------------------

%don't want date printed
\date{}

% \title{\Large \bf Formatting Submissions for a USENIX Conference:\\
%   An (Incomplete) Example}

\title{\Large \bf Seer: Online Context Learning for Fast Synchronous LLM Reinforcement Learning}

\author{
\rm Ruoyu Qin$^{\dagger\diamondsuit}$~~~~Weiran He$^{\dagger}$~~~~Weixiao Huang$^{\dagger}$~~~~Yangkun Zhang$^{\dagger}$~~~~Yikai Zhao$^{\dagger}$ \\
\rm Bo Pang$^{\dagger}$~~~~Xinran Xu$^{\dagger}$~~~~Yingdi Shan$^{\diamondsuit}$~~~~Yongwei Wu$^{\diamondsuit}$~~~~Mingxing Zhang$^{\diamondsuit1}$ \\
\rm $^{\dagger}$Moonshot AI~\medskip~$^{\diamondsuit}$Tsinghua University}

\maketitle

\footnotetext[1]{~Corresponding to zhang\_mingxing@mail.tsinghua.edu.cn.}

\begin{abstract}
Reinforcement Learning (RL) has emerged as a critical technique for advancing modern Large Language Models (LLMs), yet existing synchronous RL systems face severe performance bottlenecks. The rollout phase, which dominates end-to-end iteration time, suffers from substantial long-tail latency and poor resource utilization due to inherent workload imbalance. We present \systemname, a novel context learning RL system that addresses these challenges through a key observation: requests sharing the same prompt exhibit strong similarities in output lengths and response patterns. Leveraging this insight, \systemname introduces three coordinated techniques: (1) divided rollout for dynamic load balancing, (2) context-aware scheduling to mitigate long-tail request delays, and (3) adaptive grouped speculative decoding to accelerate generation. These mechanisms work in concert to markedly reduce long-tail latency and improve resource efficiency during rollout. Evaluations on production-grade RL workloads demonstrate that \systemname achieves up to 2.04$\times$ end-to-end rollout throughput improvement compared to the state-of-the-art synchronous RL systems, while notably reducing long-tail latency by 72--94\%.
\end{abstract}

\section{Introduction}

Reinforcement Learning (RL) has become a cornerstone in the development of state-of-the-art Large Language Models (LLMs), enabling significant breakthroughs in complex reasoning and problem-solving capabilities~\cite{team2025kimik15,guo2025deepseek,team2025kimik2}. The iterative RL training process alternates between a \emph{rollout phase} for data generation and a \emph{training phase} for updating model parameters. However, the rollout phase consistently emerges as the dominant bottleneck, consuming approximately 80\% of the total iteration time (see Table~\ref{tab:time_distribution}). 
Therefore, improving the efficiency of the rollout phase represents one of the most pressing challenges in modern LLM development.

\begin{table}[t]
\centering
\caption{Time distribution across RL training phases for different workloads. Detailed configurations given in \S\ref{sec:setup}.}
\label{tab:time_distribution}
\small
\begin{tabular}{lccc}
\toprule
 & \textbf{Rollout} & \textbf{Training} & \textbf{Weight Update} \\
\midrule
Moonlight~\cite{liu2025muon}      & 84\% & 14\% & 2\% \\
Qwen2-VL-72B~\cite{wang2024qwen2}     & 63\% & 31\% & 6\% \\
Kimi-K2~\cite{team2025kimik2}       & 87\% & 10\% & 3\% \\
\bottomrule
\end{tabular}
\end{table}

The primary challenge in RL rollout arises from severe resource inefficiency driven by the increasing demand for long-generation capabilities, especially in tasks requiring complex chain-of-thought (CoT) reasoning. These workloads create two fundamental bottlenecks. First, long-generation requests exhibit highly {\bf unpredictable and rapidly increasing memory footprints}: a CoT request may begin with only a few hundred megabytes of KVCache usage but expand to tens of gigabytes as decoding progresses. This volatility forces the system to shrink batch sizes dynamically or to preempt running requests; both outcomes reduce hardware efficiency, and preemptions are particularly costly because they trigger expensive re-prefills, ultimately degrading rollout throughput.

Second, long-generation requests produce a {\bf heavy-tailed distribution of output lengths}, resulting in pronounced load imbalance. Toward the end of a rollout iteration, only a small number of disproportionately long-running requests remain active, leaving most accelerators underutilized. The inefficiency is so severe that attempting to use idle nodes to accelerate these long-tail requests often yields little benefit and can even harm performance due to increased inter-node communication overhead.

\begin{figure}[t]
\centering
\begin{subfigure}[b]{\linewidth}
    \centering
    \includegraphics[width=\linewidth]{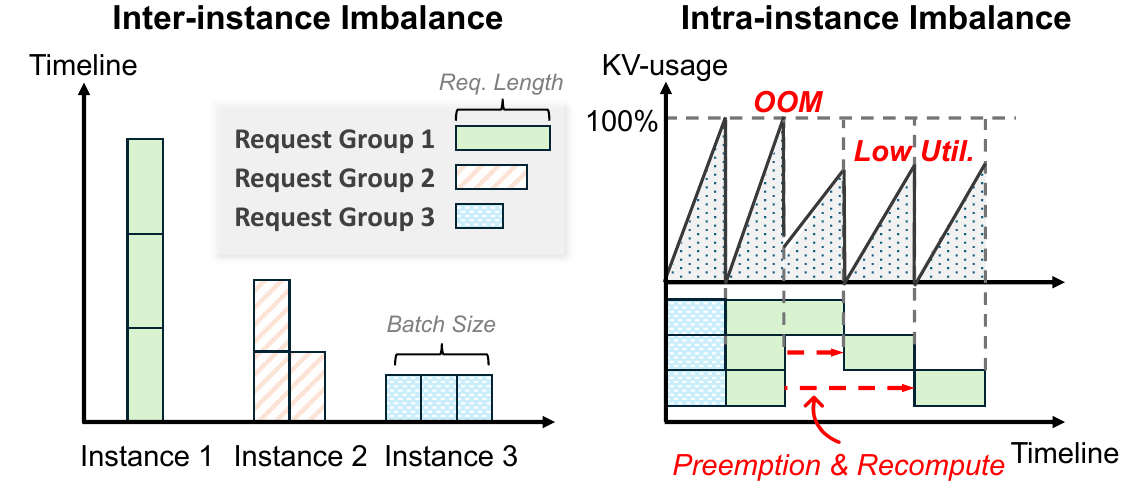}
    \caption{Group-level rollout.}
    \label{fig:baseline_rollout_illustration}
\end{subfigure}

\vspace{0.5em}

\begin{subfigure}[b]{\linewidth}
    \centering
    \includegraphics[width=\linewidth]{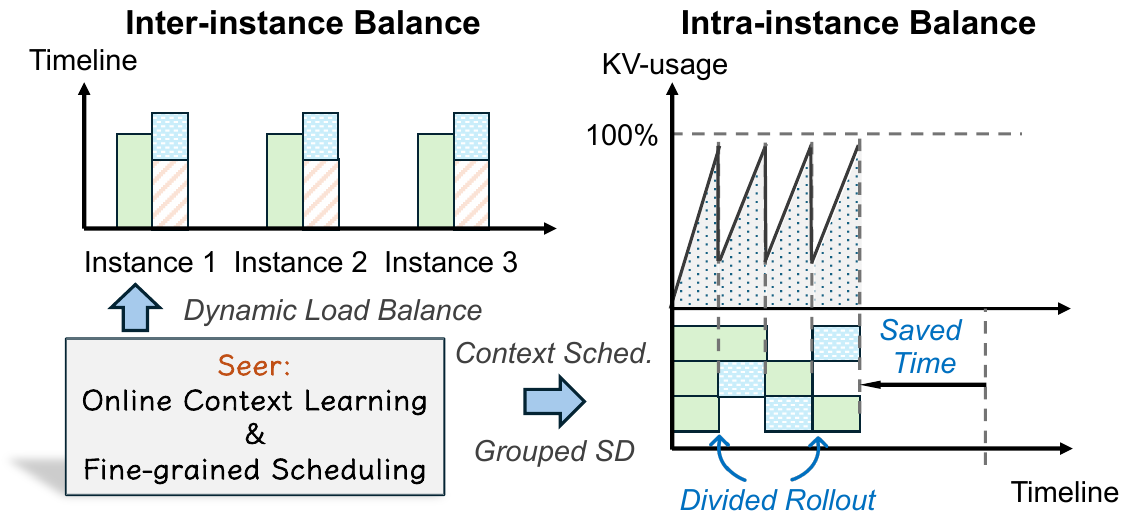}
    \caption{\systemname's rollout.}
    \label{fig:seer_rollout_illustration}
\end{subfigure}
\caption{Challenges and \systemname's solution for long-generation rollout. }
\label{fig:rollout_illustration}
\end{figure}

To improve hardware utilization, recent works have explored asynchronous rollout systems~\cite{zhong2025streamrl,fu2025areal,han2025asyncflow,sheng2025laminar,he2025history,piche2025pipelinerl}, which overlap the rollout and training phases. While this approach can reduce end-to-end iteration time, it comes at the cost of algorithmic fidelity. By nature, these systems introduce a degree of off-policy learning, as data generated with model parameters from step $i$ may be used to train the model at step $i+1$ or beyond. Furthermore, asynchronous or non-strictly synchronous~\cite{zhou2025april,gao2025rollpacker} RL systems often suffer from distributional skew, where faster-to-generate short samples disproportionately populate early training batches~\cite{meituan2025longcat}. These off-policy effects can diminish final model performance and complicate debugging and reproducibility~\cite{zheng2025stabilizing}. Consequently, synchronous (or ``on-policy'') rollout remains critical in many settings for ensuring methodical evaluation, reproducibility, and strict adherence to the underlying algorithm's assumptions. This paper, therefore, focuses on optimizing the synchronous case, although we note that our techniques can also be adapted to improve asynchronous settings.

Speculative decoding~\cite{leviathan2023fast} (SD) presents another promising direction for accelerating memory-bound generation, as its parallel verification mechanism can utilize more computational resources to accelerate individual requests. However, conventional SD methods struggle to simultaneously achieve high draft accuracy and low draft overhead in RL settings, where both the request workload and the target model itself evolve dynamically. This motivates the need for adaptive SD techniques tailored to the unique characteristics of RL rollout.

To address these challenges, we introduce \systemname, a novel system for synchronous RL rollout that dynamically exploits contextual information within the workload to maximize resource efficiency. \systemname is built upon a key observation: popular RL algorithms such as Group Relative Policy Optimization (GRPO)~\cite{shao2024deepseekmath} generate $G$ (typically 8--16) responses per prompt, and {\bf responses within a group tend to exhibit similar length profiles and recurring local token patterns}, which represent a rich structure that existing schedulers and inference engines leave untapped. For instance, generating an early ``probe'' response per prompt provides a strong, online estimator of that prompt group's remaining work (i.e., expected output length and KVCache footprint), enabling more informed scheduling decisions than static heuristics. \systemname leverages this latent intra-group context to make three key contributions:

\noindent\textit{\underline{1) Divided Rollout with Global KVCache:}} 
\systemname departs from conventional group-level scheduling, where all requests within a prompt group are treated as a single, inseparable unit. 
This traditional approach causes pronounced inter-instance and intra-instance load imbalance. As illustrated in Figure~\ref{fig:rollout_illustration}, \systemname instead performs \emph{Divided Rollout}, splitting each group not only into $G$ independent requests but further into smaller chunks that are scheduled incrementally. This fine-grained decomposition lets the scheduler pack many short requests together early in rollout to fully utilize VRAM, and later, when long requests dominate, adjust concurrency based on KVCache budgets to avoid preemptions. \systemname's global scheduler continuously monitors KVCache usage across instances and can migrate a request to a less-loaded worker when its next chunk is scheduled. This migration is efficient because \systemname uses a global KVCache pool adapted from Mooncake~\cite{qin2025mooncake} and shared across all instances, eliminating the cost of prefill recomputation.

\noindent\textit{\underline{2) Context-Aware Scheduling:}} 
\systemname leverages a \emph{``speculative request''} from each GRPO group to estimate that group’s remaining workload—specifically its likely generation length and KVCache footprint. These lightweight online estimates allow \systemname to approximate a longest-job-first scheduling policy that pairs long requests with short ones to maintain dense batches throughout rollout. Experiments in \S\ref{sec:ablation_of_context} show that this approach significantly reduces the time spent in the long-tail phase by 89\%.

\noindent\textit{\underline{3) Adaptive Grouped Speculative Decoding:}} To leverage speculative decoding for rollout acceleration while overcoming the acceptance rate collapse of traditional SD, \systemname introduces a context-learning-based speculative mechanism. \systemname deploys a Distributed Grouped Draft Server (DGDS) that maintains a Compressed Suffix Tree~\cite{weiner1973linear} (CST) for each group, aggregating token sequences from all requests within the same group. This approach creates a highly accurate, dynamic ``draft model'' that is inherently synchronized with the target model. Additionally, DGDS introduces a Marginal-Benefit-Aware Adaptive Speculation policy that balances overall throughput with the latency of high-priority requests. Our ablation experiments in \S\ref{sec:ablation_of_sd} demonstrate the effectiveness of the SD components. Our adaptive grouped speculative decoding outperforms multiple SD strategies, achieving up to 1.3$\times$ performance improvement. Compared to vanilla CST-based SD without group context, it increases the mean acceptance length by 0.22.

We have implemented \systemname and conducted an extensive evaluation on 256 H800 GPUs across multiple production-grade RL workloads, encompassing models ranging from tens of billions to over one trillion parameters. Our experiments demonstrate that \systemname consistently improves end-to-end rollout throughput by 44–104\% and cuts long-tail latency by 72–94\% relative to a highly optimized synchronous baseline. Through detailed ablations, we isolate the contributions of divided rollout, context-aware scheduling, and adaptive grouped speculative decoding, showing that each component provides substantial and complementary gains. We further compare \systemname against state-of-the-art rollout optimization approaches, including a non-strictly synchronous system~\cite{zhou2025april}, as well as multiple speculative decoding strategies~\cite{oliaro2025suffixdecoding,leviathan2023fast,liu2024deepseekv3}. We find that \systemname achieves the best overall performance while preserving strict on-policy training semantics.
\section{Background and Motivation}

\begin{figure}[t]
\centering
\includegraphics[width=\linewidth]{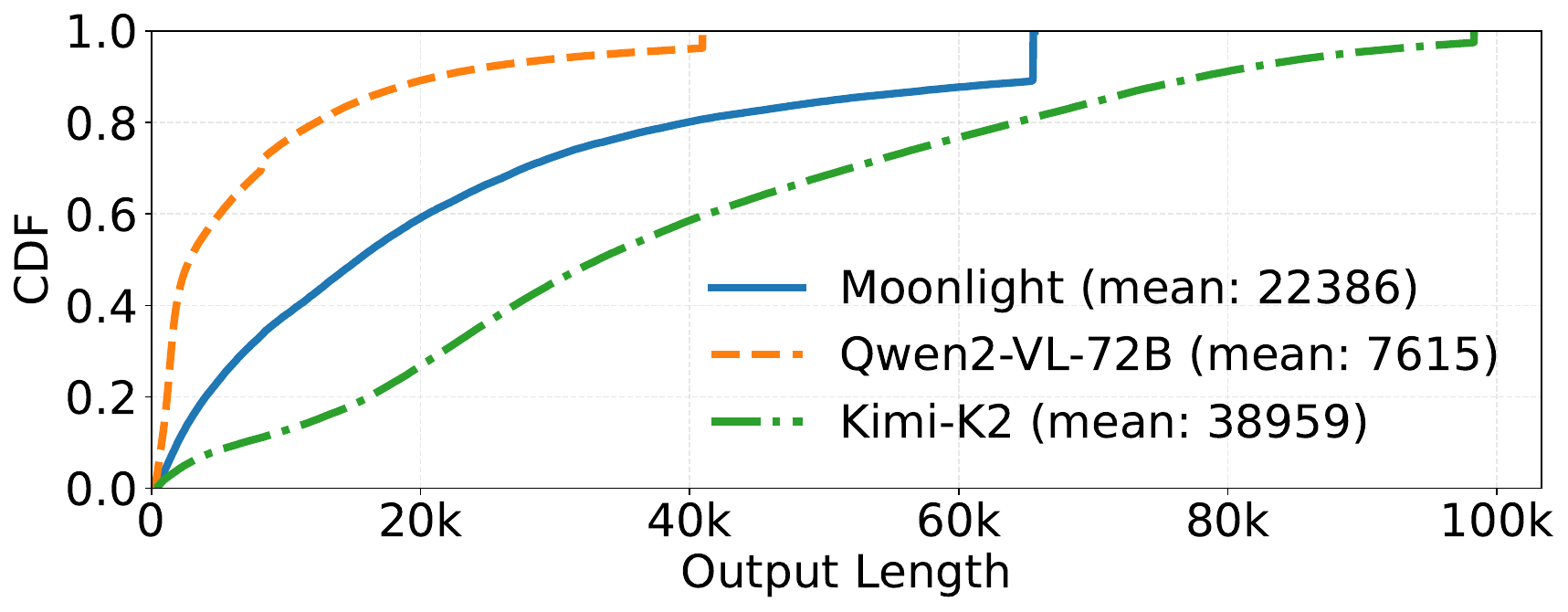}
\caption{Distribution of output lengths during rollout across three reasoning tasks. }
\label{fig:length_distribution}
\end{figure}

\begin{figure}[t]
\centering
\begin{subfigure}[b]{\linewidth}
    \centering
    \includegraphics[width=\linewidth]{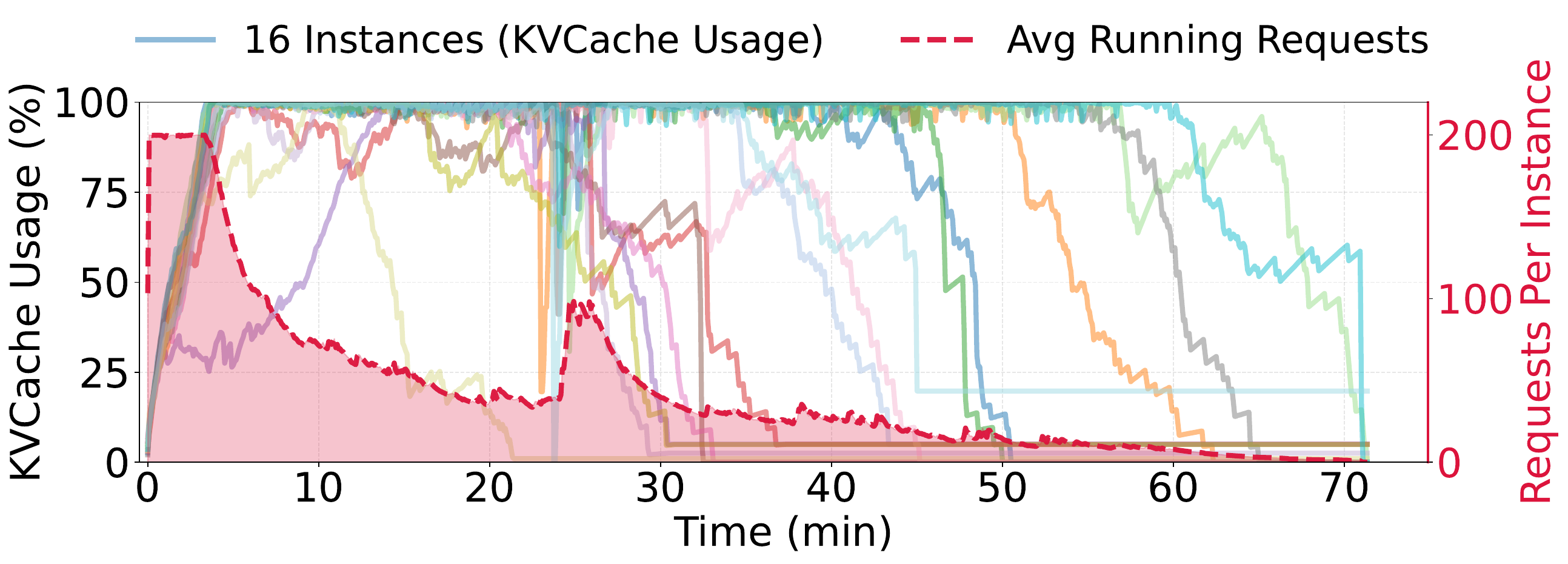}
    \caption{KVCache utilization and average number of running requests across instances.}
    \label{fig:kv_util}
\end{subfigure}

\vspace{0.5em}

\begin{subfigure}[b]{\linewidth}
    \includegraphics[width=0.92\linewidth]{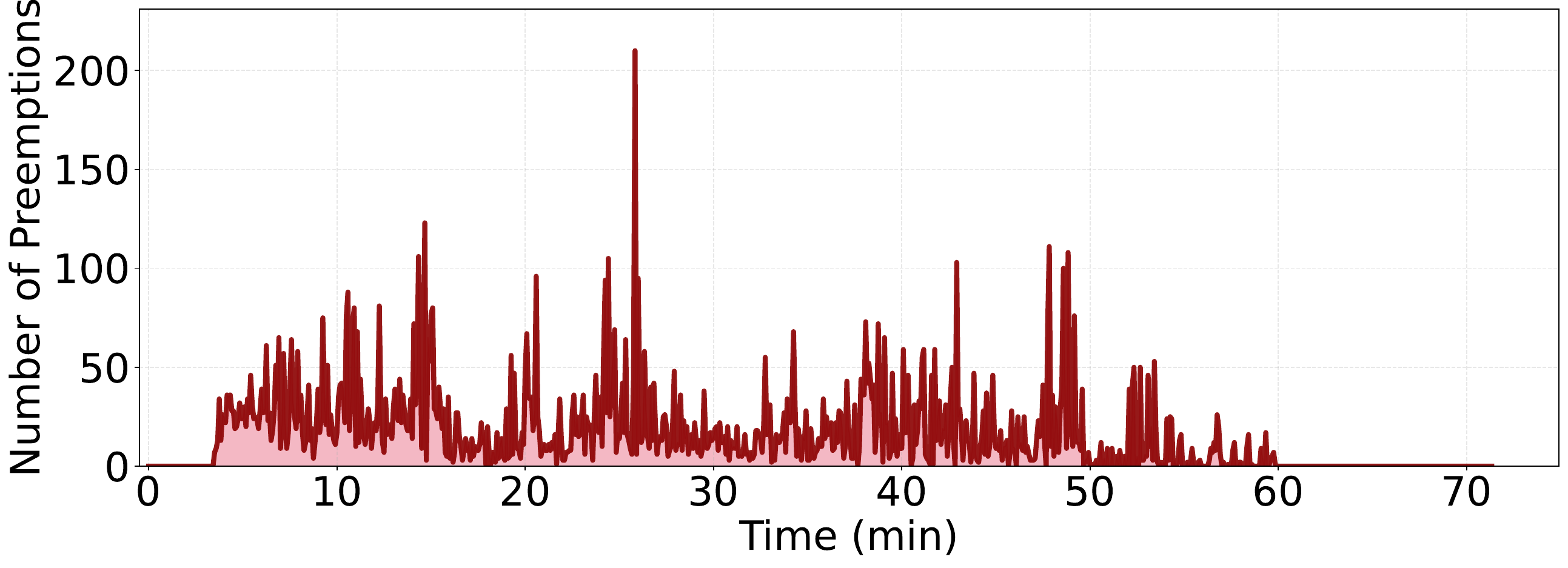}
    \caption{Request preemption count.}
    \label{fig:preempt_count}
\end{subfigure}
\caption{KVCache utilization, number of running requests, and preemption count during a synchronous rollout phase of the Qwen2-VL-72B task. In the early stage of rollout, insufficient KVCache capacity causes frequent request preemptions; in the later stage, a small number of extremely long request groups contribute to a long-tail period that accounts for nearly half of the total rollout time.}
\label{fig:rollout_analysis}
\end{figure}

Reinforcement learning (RL) for LLMs progresses through a repeated iterative loop. In each iteration, the model generates responses for a batch of prompts (rollout), evaluators assign quality scores to those responses (reward computation), these scores are transformed into supervision signals (experience construction), and the model is updated accordingly (training). The updated parameters are then distributed to inference workers to initiate the next iteration (weight update). A series of recent systems aim to accelerate this loop by improving coordination across stages. veRL~\cite{sheng2025hybridflow} colocates rollout and training to reduce data movement, RLHFuse~\cite{zhong2025optimizing} overlaps reward computation with experience construction, and Kimi-K2~\cite{team2025kimik2} further optimizes checkpoint conversion and weight distribution. 
However, Table \ref{tab:time_distribution} makes clear that such non-rollout phase optimizations are insufficient: across three representative production workloads, rollout alone occupies 63–87\% of total iteration time, overwhelmingly dominating all other stages. 

This imbalance is structural and rooted in the nature of training reasoning models. RL increasingly relies on chain-of-thought (CoT) supervision, which encourages models to produce longer and more detailed reasoning traces. As a result, rollout exhibits two defining characteristics: long average output lengths and extremely high variance across requests. Figure \ref{fig:length_distribution} shows that generations range from a few hundred tokens to as many as 96k tokens. Long average outputs exert heavy pressure on memory because KVCache consumption scales with request length, while the extreme variance produces severe long-tail effects: a small number of exceptionally long requests monopolize GPU resources near the end of rollout. Figure \ref{fig:rollout_analysis} illustrates how these properties cause substantial resource underutilization and imbalance, motivating the two challenges discussed next.

\subsection{Challenge \#1: Request Scheduling Dilemma}
\label{sec:rollout_challenge1}

During the rollout process of long-CoT models, the KVCache memory of requests undergoes dramatic changes, from negligible in the early phase to several gigabytes per request in the final stage. This varying memory consumption creates a dilemma in concurrency management. If concurrency is not controlled, the increasing request length leads to memory exhaustion, resulting in massive \textbf{request preemption}, where the KVCache of some running requests is evicted to allow a few requests to complete inference. This introduces huge overhead for KVCache recomputation. Conversely, if concurrency is controlled to avoid preemption, requests in their early generation phase (with only a few hundred tokens) may occupy the entire memory space for hundreds of seconds, leading to severe \textbf{resource underutilization}. Furthermore, this memory footprint is amplified by a factor of \(G\) in GRPO-like algorithms, where \(G\) responses are generated for each prompt within the same group, further exacerbating the inter-instance imbalance.

To address the inter-instance imbalance caused by group-level scheduling, a concurrent work Roll Flash~\cite{lu2025part} proposes \emph{prompt replication}, which splits request groups into independent requests for separate scheduling, thereby resolving the amplified imbalance caused by scheduling entire request groups. However, this does not address the dilemma of \emph{intra-instance} concurrent scheduling, and still suffers from uneven request distribution across instances leading to inter-instance imbalance. The root cause of these issues is treating individual (or groups of) requests as indivisible monolithic units, even though rollout scenarios impose no strict latency constraints on individual requests. This persistent scheduling dilemma motivates our design of \textit{divided rollout} (detailed in \S\ref{sec:divided_rollout}), which schedules requests at the chunk level with little overhead, enabling fine-grained scheduling and dynamic load balancing.

\subsection{Challenge \#2: Severe Long-tail Effect}
\label{sec:rollout_challenge2}

The long-tail problem is another critical issue in rollout, widely noted in prior work~\cite{zhou2025april,he2025history,gao2025rollpacker}. As shown in Figure~\ref{fig:kv_util}, the long-tail phase can account for nearly 50\% of the total rollout time. This phenomenon arises from two factors: (1) In GRPO-like algorithms, requests within the same group tend to have similar lengths. Groups with extremely long average lengths form ``monolithic'' batches that cause severe load imbalance across instances when scheduling is performed at the group granularity. (2) Under memory constraints, requests may be preempted or delayed, causing extremely long requests to be blocked and further exacerbating tail latency.

To mitigate the long-tail effect, recent works~\cite{zhong2025streamrl,fu2025areal,han2025asyncflow,sheng2025laminar,he2025history} have proposed \emph{asynchronous RL}. Unlike synchronous methods, where all training experiences must originate from the current policy iteration, asynchronous methods allow training on stale rollout data from previous policy iterations, introducing \emph{off-policyness}. Other works~\cite{zhou2025april,gao2025rollpacker}, while nominally maintaining synchrony, allow deferring a portion of requests to subsequent iterations, thereby sacrificing iteration-level consistency compared to strict synchronous RL systems. Although asynchronous or non-strictly synchronous RL can reduce the long-tail impact, many RL practitioners still prefer synchronous training to achieve the best convergence and maintain reward stability.

Speculative decoding (SD)~\cite{leviathan2023fast,fu2024break,li2024eagle1,li2024eagle2,li2025eagle3,cai2024medusa,liu2024deepseekv3,oliaro2025suffixdecoding,hu2025sam} offers a promising approach to accelerate the long-tail phase while preserving synchronous RL training guarantees. SD consists of two stages: \textit{Draft} and \textit{Verification}. In the draft stage, a draft model generates draft tokens, which are then verified in parallel by the target model (i.e., the policy LLM) in the verification stage. Since parallel verification of $n$ tokens by the LLM is faster than serial generation of $n$ tokens due to reduced memory access, SD can improve the generation speed for individual requests. However, RL rollout scenarios present unique challenges: batch sizes fluctuate dramatically, and the target LLM undergoes continuous updates, causing rapid ``model drift''. Existing SD strategies suffer from either high draft overhead or low draft token accuracy, resulting in marginal or even negative performance gains in rollout scenarios.

\subsection{Opportunity: Shared Contextual Information in Group Sampling}
\label{sec:group_context}\label{sec:workload_analysis}\label{sec:length_context}

In LLM reinforcement learning practice, the most widely adopted algorithm is Group Relative Policy Optimization (GRPO)~\cite{shao2024deepseekmath}. GRPO performs group-based preference optimization by sampling $G$ candidate responses for each prompt, computing rewards for these responses, and then normalizing the rewards \emph{within} the group to obtain advantages. Several recent works~\cite{yu2025dapo,zheng2025group} refine GRPO to improve generalization, but they all preserve the core principle of generating multiple responses for the same prompt. Other efforts, such as BroRL~\cite{hu2025brorl} and Knapsack RL~\cite{li2025knapsack}, even scale $G$ to 512 or more to enhance exploration. As a result, GRPO and its variants establish \emph{group sampling} as a dominant training paradigm, in which each prompt is associated with a set of responses generated by the same policy.

This group sampling paradigm naturally introduces strong and stable contextual signals during rollout. Since all responses in a group are conditioned on the same prompt and produced by the same model, they exhibit notable similarity in semantics, structural templates, and generation length. However, existing RL systems typically treat each prompt group as a monolithic unit during rollout, overlooking the potential to externalize and exploit such intra-group similarity as \textbf{shared contextual information}. Our statistical analysis shows that this shared context is sufficiently rich to support both scheduling and inference optimizations.

\begin{figure}[t]
\centering
\includegraphics[width=\linewidth]{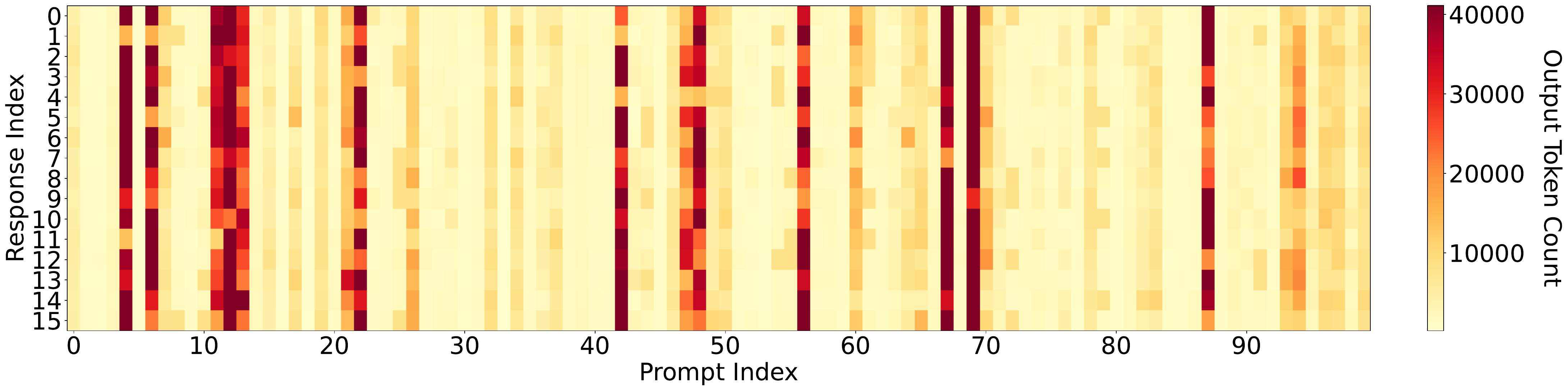}
\caption{Length correlation within response groups. Each column represents a prompt group in GRPO rollout, and each cell corresponds to an individual response. The color intensity indicates output length.}
\label{fig:length_correlation}
\end{figure}

First, at the \emph{length} level, responses within the same group have highly correlated generation lengths. Prior work~\cite{cheng2024enabling,fu2024efficient,zhong2025streamrl} has shown that response length is predictable from prompt and model characteristics. Our measurements on real RL workloads, summarized in Figure~\ref{fig:length_correlation}, further confirm that most prompt groups in GRPO rollout exhibit strong length correlation. We observe that responses within the same group tend to have similar lengths, forming visually consistent columns. This length context can be obtained cheaply via speculative sampling and used to inform global scheduling policies, such as approximate longest-first scheduling, which prioritizes groups likely to produce very long responses. Such length- and memory-aware scheduling is a promising direction for mitigating the long-tail effect while respecting device memory constraints.

\begin{table}[t]
\centering
\caption{Mean acceptance length in n-gram speculative decoding with grouped pattern references under different draft strategies.  Linear generates one draft sequence per step, while multi-path generates multiple candidate sequences with top-$k$ branching. Values represent mean acceptance length (including bonus token).}
\label{tab:pattern_similarity}
\small
\begin{tabular}{@{}lccc@{}}
\toprule
\textbf{Ref. Count} & \textbf{Linear} & \textbf{Multi-Path (k=2)} & \textbf{Multi-Path (k=4)} \\
\midrule
$n=0$ (baseline) & 1.70 & 1.77 & 1.85 \\
$n=1$ & 2.04 & 2.14 & 2.25 \\
$n=5$ & 2.32 & 2.44 & 2.59 \\
$n=15$ & 2.53 & 2.69 & 2.85 \\
\bottomrule
\end{tabular}
\end{table}

Second, at the \emph{pattern} level, grouped responses share recurring semantic and syntactic structures that can be exploited to accelerate inference. Instead of relying solely on each request’s own history, we can aggregate previously generated tokens from other responses in the same group into a shared pattern dictionary (e.g., via compressed suffix trees) and use it as an n-gram reference for speculative decoding. To quantify this opportunity, we sample 20 prompt groups from a Qwen2-VL-72B task and simulate n-gram speculative decoding under different draft strategies. Table~\ref{tab:pattern_similarity} reports the mean acceptance length (including bonus tokens) for varying numbers of grouped references $n$ and different drafting modes. Compared to the self-referencing baseline with no grouped references ($n=0$), incorporating even a small number of intra-group references (e.g., $n=1$ or $n=5$) consistently increases the mean acceptance length, and using more references (e.g., $n=15$) together with multi-path drafting yields the largest gains. 
In the later stage of rollout, speculative decoding with full grouped pattern references can improve the number of accepted draft tokens by up to 119\% compared to a baseline that only uses per-request history.

In summary, group sampling in GRPO-like algorithms provides rich shared contextual information, both in terms of length correlation and pattern similarity, that is currently underutilized by existing RL systems. Our statistical data indicates that this contextual signal can be harnessed to enable length- and memory-aware scheduling and grouped speculative decoding, offering a promising opportunity to mitigate long-tail inefficiency and improve rollout throughput in synchronous LLM RL training.
\section{Design of \systemname}

\begin{figure}[t]
\begin{center}
    \includegraphics[width=\linewidth]{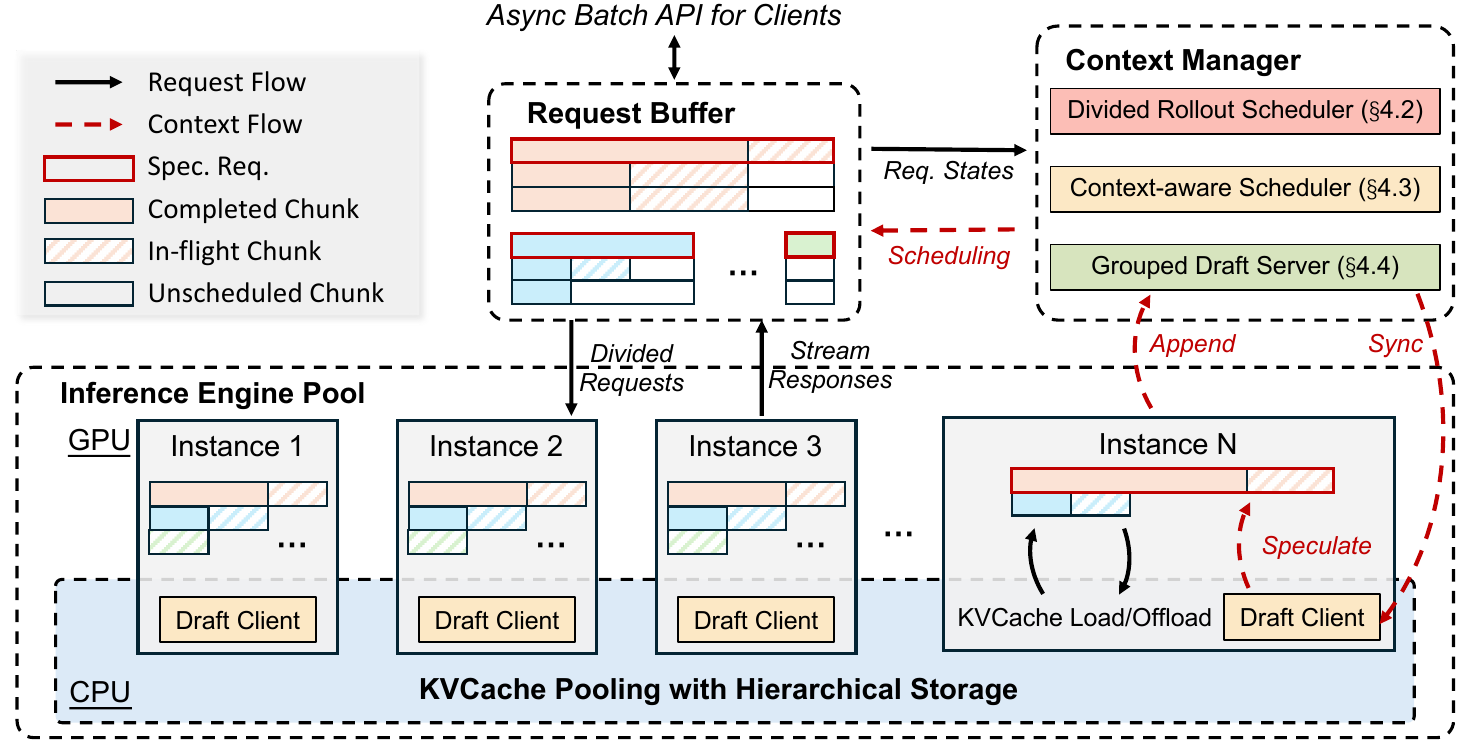}
    \caption{The overview of \systemname.}
    \label{fig:seer_architecture}
\end{center}
\end{figure}

\subsection{Overview}

\systemname is a synchronous, colocated RL system designed to substantially reduce rollout long-tail latency and improve overall system throughput without compromising algorithmic fidelity. The broader pipeline remains standard: training uses Megatron~\cite{shoeybi2019megatron} for distributed optimization, rollout uses vLLM~\cite{kwon2023efficient} for generation and an asynchronous reward computation backend, and Moonshot Checkpoint Engine~\cite{checkpoint2025} ensures rapid movement of updated model weights. \systemname's contributions therefore layer onto a conventional architecture without imposing additional integration overhead.

Within rollout, as motivated by the workload analysis in \S\ref{sec:workload_analysis}, \systemname introduces a unifying idea that drives its three innovations: Group-Aware Context Learning. \systemname continuously learns the intra-group shared properties and uses them to guide scheduling and decoding. Length similarities become signals for estimating whether upcoming generations are likely to be long-running; pattern similarities reveal opportunities to construct group-level draft predictions that accelerate decoding without a separate draft model. In addition to these optimizations, \systemname also incorporates efficient memory management, load balancing, and asynchronous reward computation. 

To support this context-driven approach at scale, the rollout subsystem is organized into three tightly connected components as illustrated in Figure~\ref{fig:seer_architecture}. An \texttt{Inference Engine Pool} hosts multiple model-serving instances backed by a distributed KVCache pool, enabling KVCache migration across nodes without recomputation. A \texttt{Request Buffer} provides a global view of all pending and in-progress rollout work, allowing the system to coordinate concurrency precisely. Atop these components, a logically centralized \texttt{Context Manager} maintains group-level contextual views and drives scheduling and draft generation decisions.

On this foundation, \systemname introduces three techniques that address the underlying challenges of long-CoT rollout. Divided Rollout (\S\ref{sec:divided_rollout}) resolves the concurrency–memory dilemma by breaking each request into smaller schedulable units and enabling fine-grained placement across instances. Context-Aware Scheduling  (\S\ref{sec:context_aware_scheduling}) then leverages learned length context to approximate a longest-first policy that reduces tail latency. Finally, Adaptive Grouped Speculative Decoding (\S\ref{sec:speculative_decoding}) exploits shared token-pattern structure to build speculative decoding at the group level and dynamically adjusts draft lengths to maximize throughput. Despite these performance enhancements, \systemname maintains logical consistency across all stages of the synchronous RL pipeline, thereby achieving an algorithmically lossless reinforcement learning process.

\subsection{Divided Rollout}
\label{sec:divided_rollout}

As illustrated in Figure~\ref{fig:rollout_illustration}, conventional group-level request scheduling mechanisms lead to severe load imbalance both within and across inference instances. 
Divided Rollout addresses this core tension by rethinking what constitutes a schedulable unit. Rather than binding any request group to a single execution slot, \systemname not only decomposes it into individual requests but further divides each request into a sequence of generation chunks, each representing a bounded segment of progress. A request that would previously monopolize an instance for thousands of tokens becomes a pipeline of shorter units that can be interleaved and redistributed as the system's load evolves. This change in granularity enables continuous rebalancing: after each chunk completes, the next chunk can be scheduled on whichever instance currently has the most available memory and compute.

However, divided rollout poses significant challenges to the KVCache system. When requests are rescheduled at the chunk level, all KVCache will be recomputed, and this overhead can even negate the benefits of load balancing. Although existing inference frameworks~\cite{vllm2025,sglang2025} support offloading request KVCache to DRAM within instances, inference instances can generate tens of terabytes of KVCache during a single rollout iteration, far exceeding the DRAM capacity of individual instances. Furthermore, due to the dynamic load balancing scheduling policy, requests may be rescheduled to other instances, leading to cache misses. Overall, divided rollout requires a globally shared, high-capacity cache system to support chunk-level request scheduling with minimal overhead.

To address this challenge, \systemname builds upon Mooncake~\cite{qin2025mooncake} to construct a globally shared KVCache pool distributed across inference nodes. The KVCache of all active requests is stored in a hierarchical global storage spanning DRAM and SSD, with RDMA enabling rapid KVCache transfer between nodes. From the perspective of the upper-level scheduling system, request chunk scheduling can be treated as stateless, eliminating the need to consider KVCache distribution.

What distinguishes Divided Rollout from prior offloading or preemption-based approaches is that cache movement is no longer a reactive measure taken only when memory pressure becomes intolerable. Instead, since chunk-level work units have well-defined memory footprints, the scheduler can proactively place and move them to optimize global system balance rather than merely avoiding out-of-memory events. This shift from reactive eviction to proactive, mobility-aware scheduling is the key novelty that allows \systemname to resolve the concurrency–memory dilemma that has long constrained synchronous RL rollout. Compared with live migration systems for online serving (e.g., Llumnix~\cite{sun2024llumnix}), it offers greater scheduling flexibility due to the absence of strict per-request latency constraints.

\subsection{Context-Aware Scheduling}
\label{sec:context_aware_scheduling}

Long-CoT rollout inevitably runs into memory limits, which means some requests must wait. When the scheduler cannot anticipate which requests will run long, these delays occur arbitrarily, creating unnecessary congestion in the tail. As shown in \S\ref{sec:length_context}, requests within the same GRPO group tend to exhibit similar output lengths. \systemname leverages this structure to predict which requests are likely to be long-running and schedules them more intelligently. The goal is to approximate a longest-first scheduling (LFS) policy (known to reduce tail latency) without needing to know true output lengths in advance.

\systemname's key idea is to designate one request per group as a \textbf{speculative request}. This request serves as an online probe: by observing how quickly it completes during generation, \systemname can infer the approximate length of the other requests in the group. To surface these signals early, speculative requests are placed in a high-priority path and scheduled according to a shortest-first strategy (SFS). Because short requests complete quickly while long ones linger, this ``length filtering'' step exposes potential long-tail groups early in the rollout, giving \systemname time to react before those long generations dominate system resources.

Specifically, the scheduling process unfolds in three conceptual phases. First, speculative requests are executed promptly so that length information can be gathered with minimal delay. Second, the Context Manager maintains and continually updates an estimated output length for each group. This estimate is simply the maximum generation length observed among completed requests in the group, an approach that naturally converges toward the true length as more information becomes available. For groups in which no request has finished yet, \systemname conservatively assumes they may be long-tail cases and initializes their estimate to the upper bound on generation length. Third, with group-level length estimates in place, the scheduler shifts to an approximate LFS policy: groups predicted to be long are prioritized so that their requests begin making progress early, rather than being deferred until system load is already high. The detailed scheduling algorithm is provided in the appendix.

\systemname also incorporates safeguards to mitigate inaccuracies in early predictions. It occasionally schedules requests from underserved groups to avoid starvation and updates group length estimates conservatively to reflect the longest sample observed so far. These mechanisms ensure stability even when speculative signals are noisy. As shown later in \S\ref{sec:ablation_of_context}, this context-aware scheduling policy approaches the performance of an oracle LFS scheduler that has perfect knowledge of request lengths. 

\subsection{Adaptive Grouped Speculative Decoding}
\label{sec:speculative_decoding}

\begin{figure}
\begin{center}
    \includegraphics[width=\linewidth]{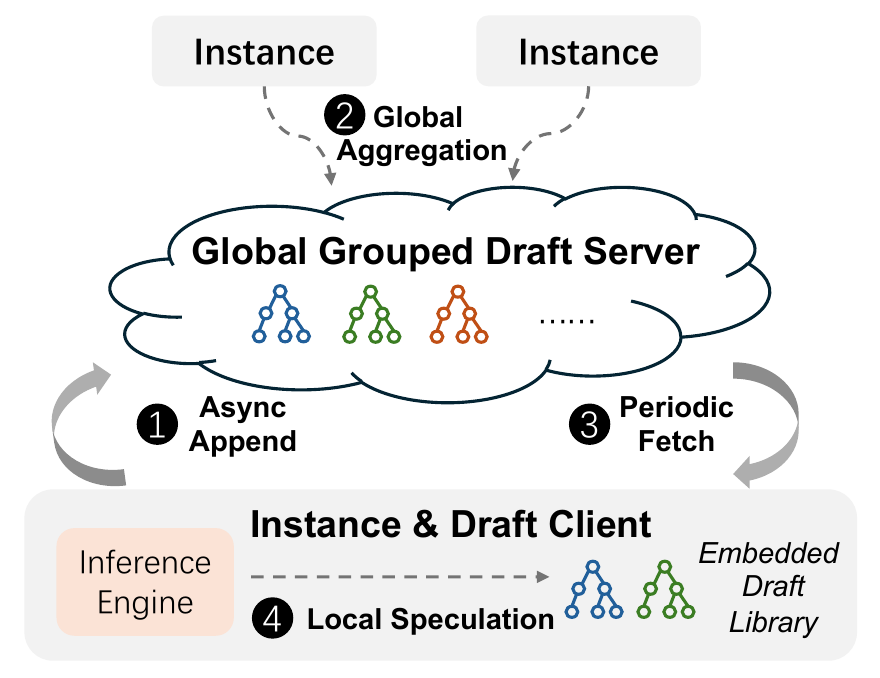}
    \caption{The distributed grouped draft server.}
    \label{fig:draft_server}
\end{center}
\end{figure}

To further improve resource utilization during the rollout phase, especially in the long-tail stage, \systemname implements Adaptive Grouped Speculative Decoding, which \textbf{adaptively} adjusts draft lengths based on computational intensity and exploits the \textbf{grouped} pattern context (as analyzed in Table~\ref{tab:pattern_similarity}) to enhance acceptance rates.

\subsubsection{Challenges of Speculative Decoding in Rollout}

Speculative decoding (SD) benefits from underutilized computational resources when batch sizes are small, enabling faster parallel verification compared to serial generation. However, rollout scenarios feature dynamically varying batch sizes that start large and rapidly decrease as long requests dominate. Fixed-length SD methods often incur excessive draft or verification overhead, degrading overall throughput. We now model SD throughput~\cite{leviathan2023fast} in the rollout scenario to identify the key factors affecting performance. Let $\alpha$ denote the acceptance rate, where $\alpha=E(\beta)$ and $\beta$ is the acceptance probability at each position. Let $\gamma$ represent the number of draft tokens predicted by the draft model per step, $B$ denote the current batch size, $D(B, \gamma)$ represent the forward time of the draft model, and $T(B, \gamma)$ represent the forward time of the target model.

The expected number of tokens generated per request in each forward is $\frac{1-\alpha^{\gamma+1}}{1-\alpha}$. The expected time for SD to generate one token per request is:
\begin{align*}
    T_{SD}=\frac{(1-\alpha)(D(B, \gamma)+T(B, \gamma))}{1-\alpha^{\gamma+1}}
\end{align*}
SD provides a benefit when $T_{SD}$ is less than the forward time of the target model $T(B, 1)$.

When $B$ is small, $D(B, \gamma)+T(B, \gamma)$ is approximately equal to $T(B, 1)$, making SD clearly beneficial. However, when $B$ is large, $D(B, \gamma)$ becomes non-negligible and $T(B, \gamma)$ increases rapidly with $\gamma$ as the target model becomes compute-bound, potentially resulting in negative gains from SD. In rollout scenarios, $B$ dynamically varies and can range from 1 to several hundred. Conventional static SD strategies with fixed draft lengths suffer from either underutilization of computational resources with small $\gamma$ or excessive verification overhead with large $\gamma$, resulting in marginal or even negative performance gains. Furthermore, to optimize SD performance, the draft system should minimize $D(B, \gamma)$ while maximizing $\alpha$. In rollout scenarios dominated by large batches, conventional SD methods suffer from either high $D(B, \gamma)$ (e.g., using a separate draft model) or low $\alpha$ (e.g., naive n-gram methods). Therefore, an efficient draft system specifically tailored to the characteristics of rollout workloads is required.

\subsubsection{Adaptive Speculation with Disaggregated Architecture}
\label{sec:sd_design}

\systemname introduces the Distributed Grouped Draft Server (DGDS), a disaggregated SD framework specifically designed for rollout scenarios. As illustrated in Figure~\ref{fig:draft_server}, DGDS aggregates response pattern context across requests and instances through an independent draft server, which asynchronously distributes the context to the embedded draft client within each inference instance. The core data structure of DGDS is the Compressed Suffix Tree (CST), which enables sharing pattern information across multiple sequences and provides draft tokens with low complexity\footnote{The complexity is $O(p+s)$, where $p$ denotes the matching pattern length and $s$ denotes the number of speculative tokens.}. Unlike previous suffix tree methods~\cite{oliaro2025suffixdecoding,vllm2025} that serialize CST updates with model execution, which increases the draft time $D(B, \gamma)$, DGDS employs a distributed master-worker architecture and adopts asynchronous CST updates to minimize speculative decoding latency in the critical path. The detailed workflow is provided in the appendix.

\begin{algorithm}[t]
\small
\caption{Marginal-Benefit-Aware Adaptive Speculation}\label{alg:adaptive_speculation}
\begin{algorithmic}[1]
\Require {
    High-priority and low-priority batch sizes $B_h$ and $B_l$;\;
    per-position acceptance probabilities $\beta[1], \beta[2], \ldots$;\;
    maximum token budget per request $\gamma_{\max}$;\;
    priority factor $\lambda \in [1, \infty)$ (e.g., $\lambda = 2$).
}
\Ensure Draft token counts $\gamma_h$ and $\gamma_l$.

\State $B \gets B_h + B_l$
\State $\gamma^{*} \gets \argmin_\gamma T_{SD}(B, \gamma)$ \Comment{optimal draft length for batch size $B$}
\State $\Gamma^{*} \gets \gamma^{*} \cdot B$ \Comment{total token budget}

\If{$\Gamma^{*} < B_h$}
    \State \Return $(\gamma_h = 0, \gamma_l = 0)$ \Comment{disable speculation}
\EndIf

\State \Comment{Allocate budget based on marginal benefit}
\State $\gamma_h \gets 1$, $\gamma_l \gets 0, \mathit{remaining} \gets \Gamma^{*} - B_h$

\While{$\mathit{remaining} > 0$}
    \State $\mathit{benefit}_h \gets B_h \cdot (\beta[\gamma_h]-\beta[\gamma_h + 1]) $
    \State $\mathit{benefit}_l \gets B_l \cdot (\beta[\gamma_l]-\beta[\gamma_l + 1])$
    
    \If{$\mathit{benefit}_h > \lambda \cdot \mathit{benefit}_l$ \textbf{and} $\gamma_h < \gamma_{\max}$ \textbf{and} $\mathit{remaining} \geq B_h$}
        \State $\gamma_h \gets \gamma_h + 1$ \Comment{allocate to high-priority}
        \State $\mathit{remaining} \gets \mathit{remaining} - B_h$
    \ElsIf{$B_l > 0$ \textbf{and} $\gamma_l < \gamma_{\max}$ \textbf{and} $\mathit{remaining} \geq B_l$}
        \State $\gamma_l \gets \gamma_l + 1$ \Comment{allocate to low-priority}
        \State $\mathit{remaining} \gets \mathit{remaining} - B_l$
    \Else
        \State \textbf{break} \Comment{cannot allocate further}
    \EndIf
\EndWhile

\State \Return $(\gamma_h, \gamma_l)$
\end{algorithmic}
\end{algorithm}

To ensure that SD consistently provides positive gains throughout the rollout process, it is necessary to dynamically adjust $\gamma$ based on the current batch size according to the throughput model. Furthermore, as described in \S\ref{sec:context_aware_scheduling}, requests are classified into high-priority and low-priority categories. High-priority speculative requests are used to probe the length distribution of requests within the same group and should complete faster, thus requiring higher draft budgets. However, acceptance rates decrease rapidly with increasing draft length, making overly disparate budget allocations between high-priority and low-priority requests wasteful. To address this challenge, we propose a \textbf{Marginal-Benefit-Aware (MBA) Adaptive Speculation} policy inspired by classical utility maximization and marginal-utility scheduling principles widely used in resource allocation systems~\cite{kelly1998rate}, which dynamically balances overall throughput with the latency of high-priority requests.

Algorithm~\ref{alg:adaptive_speculation} presents the MBA strategy. Using offline-profiled $T_{SD}$ models and online-collected acceptance rates and batch sizes, the algorithm determines draft lengths $(\gamma_h, \gamma_l)$ for high-priority and low-priority requests. The algorithm is invoked periodically during rollout, as request characteristics remain relatively stable over short time intervals. Given the draft lengths, the embedded draft client in DGDS generates the corresponding number of draft tokens based on the grouped CSTs for each request. The basic single-path speculation algorithm follows SuffixDecoding~\cite{oliaro2025suffixdecoding}, where each candidate path is assigned a confidence score computed from suffix probabilities. Building on this foundation, \systemname uses these scores to filter low-probability candidates and is capable of returning multiple candidate paths via a beam-search mechanism.

For long-tail requests, adaptive grouped speculative decoding offers two particular advantages. First, during the long-tail stage, concurrency is minimal, allowing larger draft lengths to increase the number of accepted tokens per request. \systemname also implements multi-path speculative decoding to further improve the acceptance length in the long-tail stage. Second, as more requests in the same group complete over time, the CST aggregates richer contextual information. As demonstrated in Table~\ref{tab:pattern_similarity}, this enables long-tail requests to achieve substantially longer acceptance lengths.
\section{Evaluation}
\label{sec:evaluation}

In this section, we evaluate \systemname's performance advantages during the rollout phase, particularly in improving throughput and reducing tail latency (\S\ref{sec:e2e}). In the ablation study (\S\ref{sec:ablation}), we provide a detailed analysis of how each technique in \systemname contributes to improving end-to-end performance. Finally, we present extended studies (\S\ref{sec:case_study}) on scheduling, speculative decoding, and non-strictly synchronous RL to demonstrate the effectiveness of our context-aware techniques.

\subsection{Setup}
\label{sec:setup}

\begin{table}[t]
\centering
\caption{Model configurations and RL workload characteristics.}
\label{tab:model_config}
\small
\begin{tabular}{lccc}
\toprule
\textbf{Metric} & \textbf{Moonlight} & \textbf{Qwen2-VL-72B} & \textbf{Kimi-K2} \\
\midrule
Model Size & 32 GB & 146 GB & 1 TB \\
Total GPUs & 32 & 128 & 256 \\
GPUs per Instance & 1 & 8 & 32 \\
Reqs per Iter & 3200 & 9600 & 6400 \\
Group Size & 8 & 16 & 8 \\
Temperature & 1.0 & 0.8 & 1.0 \\
Max. Gen. Length & 65536 & 40960 & 98304 \\
Avg. Gen. Length & 22386 & 7615 & 38959 \\
\bottomrule
\end{tabular}
\end{table}

\paragraph{Testbed.}
Our experimental infrastructure consists of 32 high-performance compute nodes, each equipped with 8$\times$H800 GPUs, 224 CPU cores, 2TB DRAM, and 4TB NVMe storage, providing sufficient hardware support for the global KVCache pool and distributed grouped draft server. For deployment, we adopt a strategy that balances task performance with resource efficiency by configuring different numbers of GPUs and parallelization strategies to serve models based on their size and architecture.

\paragraph{Models and Workloads.}
To validate the generalizability of \systemname's design, we evaluate three models with diverse sizes and output characteristics: Moonlight~\cite{liu2025muon}, Qwen2-VL-72B~\cite{wang2024qwen2}, and Kimi-K2~\cite{team2025kimik2}. All models are trained as \textbf{reasoning models} with chain-of-thought capabilities, where Moonlight and Kimi-K2 are trained on mathematical datasets, while Qwen2-VL-72B is trained on language-vision-mixed reasoning tasks using the LLM-as-a-Judge~\cite{son2024llm} reward model. Their configurations and RL workload characteristics are shown in Table~\ref{tab:model_config}. We employ Qwen2-VL-72B with tensor parallelism (TP8) and Kimi-K2 with data parallelism (DP32) and expert parallelism (EP32).

\paragraph{Settings.}
We use the GRPO~\cite{shao2024deepseekmath} algorithm in our experiments. The group size (i.e., the number of requests per prompt) follows RL training conventions unless otherwise specified: we use a smaller group size (8) for mathematical problems and a larger group size (16) for open-ended tasks with LLM-as-a-Judge evaluation. For \systemname's SD hyperparameters described in \S\ref{sec:sd_design}, we set $\gamma_{max} = 8$ and $\lambda = 2$.

\paragraph{Baselines.}
We evaluate \systemname against a set of strong and representative baselines that capture the best available techniques for synchronous rollout scheduling, skewness mitigation, and speculative decoding.

\begin{enumerate}[label=(\arabic*),topsep=0pt,itemsep=0pt,parsep=0pt,leftmargin=20pt]
    \item \emph{veRL}~\cite{sheng2025hybridflow}: veRL is a state-of-the-art synchronous RL system that supports efficient colocation of training and rollout. It provides a well-engineered baseline for measuring \systemname's end-to-end improvements because it already incorporates optimized model execution pipelines and a production-grade rollout subsystem.
    \item \emph{StreamRL-Oracle}: Although StreamRL \cite{zhong2025streamrl} is designed as a disaggregated asynchronous RL framework, its proposed Skewness-Aware Scheduling directly targets the same long-tail latency issues that appear in synchronous rollout. StreamRL’s scheduling relies on a small auxiliary model that predicts prompt lengths and uses those predictions to perform bucketing and LFS-style scheduling. To ensure a {\bf fair and stringent comparison}, we evaluate against StreamRL-Oracle, which uses the ground-truth prompt lengths obtained from veRL’s rollout traces rather than model predictions. This removes the confounding effects of prediction errors and reflects the best-case performance achievable by the StreamRL scheduling approach.
    \item \emph{Vanilla Speculative Decoding (SD)}: To benchmark against modern decoding accelerators, we implement strong speculative decoding baselines tailored to each model family. For Moonlight, we use SuffixDecoding \cite{oliaro2025suffixdecoding} with a maximum draft length of $\gamma_{max}=16$. For Qwen2-VL-72B, we adopt a dedicated Qwen2-7B-VL draft model with $\gamma_{max}=3$. For Kimi-K2, we apply Multi-Token Prediction (MTP) \cite{liu2024deepseekv3} with $\gamma_{max}=1$. Each of these SD methods is integrated into both veRL and StreamRL-Oracle pipelines, enabling direct comparisons across scheduling regimes and decoding strategies.
\end{enumerate}
To isolate the effect of scheduling and speculative mechanisms, all baselines (including \systemname) use a unified in-house implementation of vLLM \cite{kwon2023efficient} as the inference engine. This ensures that differences in performance arise solely 
from the scheduling and decoding techniques under evaluation, not from underlying infrastructure discrepancies.

\paragraph{Metrics.}
For the end-to-end experiments, we measure the average rollout throughput across 5 iterations, which is defined as the average number of output tokens generated per second in each rollout iteration.

\subsection{End-to-End Performance}
\label{sec:e2e}

\begin{figure*}[t]
\centering
\includegraphics[width=\textwidth]{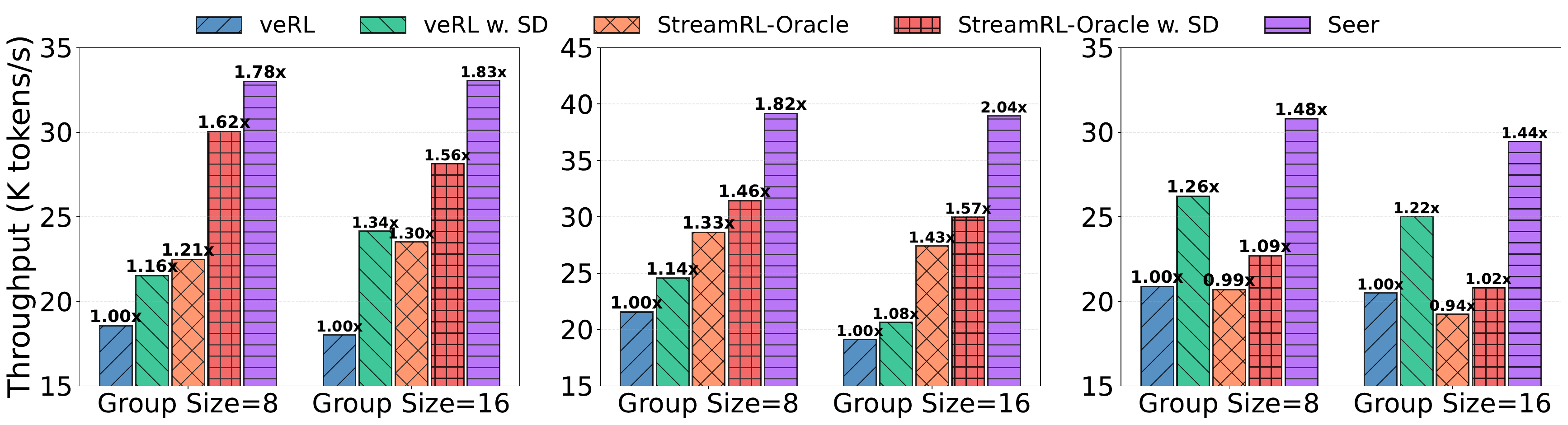}\\[0.5em]
\begin{minipage}[t]{0.33\textwidth}
    \centering
    (a) Moonlight
\end{minipage}%
\begin{minipage}[t]{0.33\textwidth}
    \centering
    (b) Qwen2-VL-72B
\end{minipage}%
\begin{minipage}[t]{0.33\textwidth}
    \centering
    (c) Kimi-K2
\end{minipage}
\caption{End-to-end rollout throughput of RL systems across different tasks and group sizes.}
\label{fig:e2e_throughput}
\vspace{-1em}
\end{figure*}

In our end-to-end experiments, we conduct three RL tasks based on the aforementioned models, with workload configurations detailed in Table~\ref{tab:model_config}. We provide a comprehensive comparison of \systemname's end-to-end performance against different baselines in \S\ref{sec:rollout_throughput}. To validate \systemname's effectiveness in mitigating tail latency, we analyze the tail latency phenomenon in our experiments and compare the tail latency between \systemname and the baseline system veRL in \S\ref{sec:tail_latency}.

\subsubsection{Rollout Throughput}
\label{sec:rollout_throughput}

We compare the throughput performance of \systemname against state-of-the-art baselines across different workloads and group sizes, as shown in Figure~\ref{fig:e2e_throughput}. Despite significant variations in model size and workload characteristics across different RL tasks, \systemname consistently achieves substantial speedups across all tasks, with throughput improvements ranging from 44\% to 104\% over veRL. 

\systemname also outperforms StreamRL-Oracle and multiple SD strategies. StreamRL sets buckets with smaller concurrency for long-request groups to minimize latency. However, it still treats each request group as an atomic, non-preemptible unit and cannot dynamically adjust instance loads at runtime. Consequently, its performance is heavily dependent on the accuracy of the preset bucketing algorithm. When encountering out-of-distribution workloads such as Kimi-K2, StreamRL-Oracle even underperforms veRL, which uses a simple round-robin scheduling strategy. For vanilla SD methods, although we have implemented adaptive draft length, these SD methods still suffer from either excessive overhead or low average acceptance length, resulting in inferior performance compared to our adaptive grouped speculative decoding. A detailed comparison is provided in \S\ref{sec:ablation_of_sd}.

Across different workloads, \systemname achieves greater performance improvements on memory-constrained tasks (Moonlight and Qwen2-VL-72B), where \systemname's divided rollout mitigates load imbalance and context-aware scheduling reduces long request delays. Furthermore, as the group size increases, the load imbalance caused by veRL's group-level scheduling becomes more severe, leading to throughput degradation. In contrast, \systemname addresses the monolithic request group problem through divided rollout and dynamic load balancing, and leverages group context to optimize both scheduling and speculative decoding. As a result, when the group size increases from 8 to 16, \systemname achieves an average performance improvement of 5\%.

\subsubsection{Long-Tail Time}
\label{sec:tail_latency}

\begin{figure}[t]
\centering
\includegraphics[width=\columnwidth]{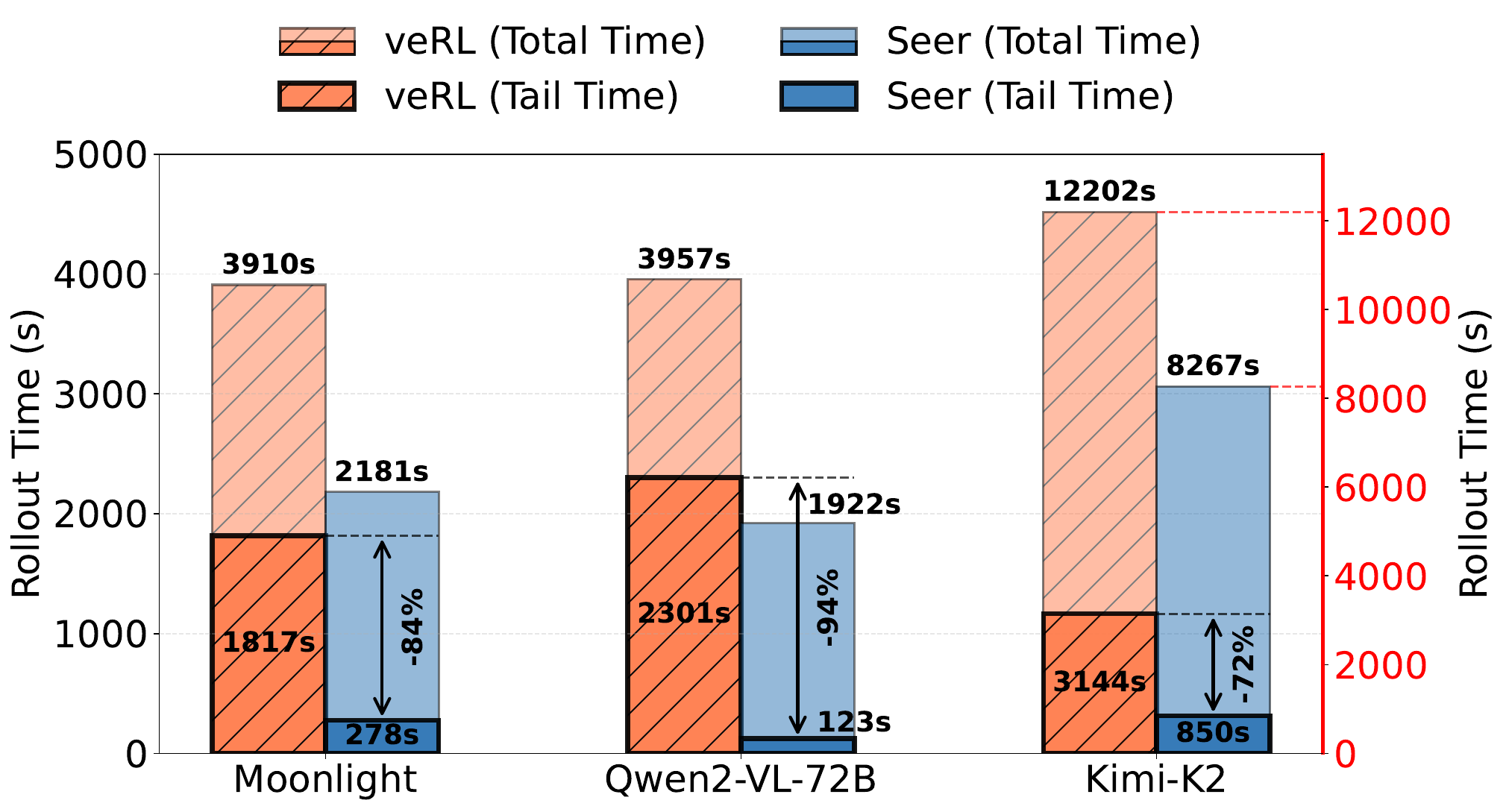}
\caption{Tail time and total time of three RL tasks.}
\label{fig:tail_latency}
\vspace{-0.5em}
\end{figure}

\begin{figure}[t]
\centering
\includegraphics[width=\columnwidth]{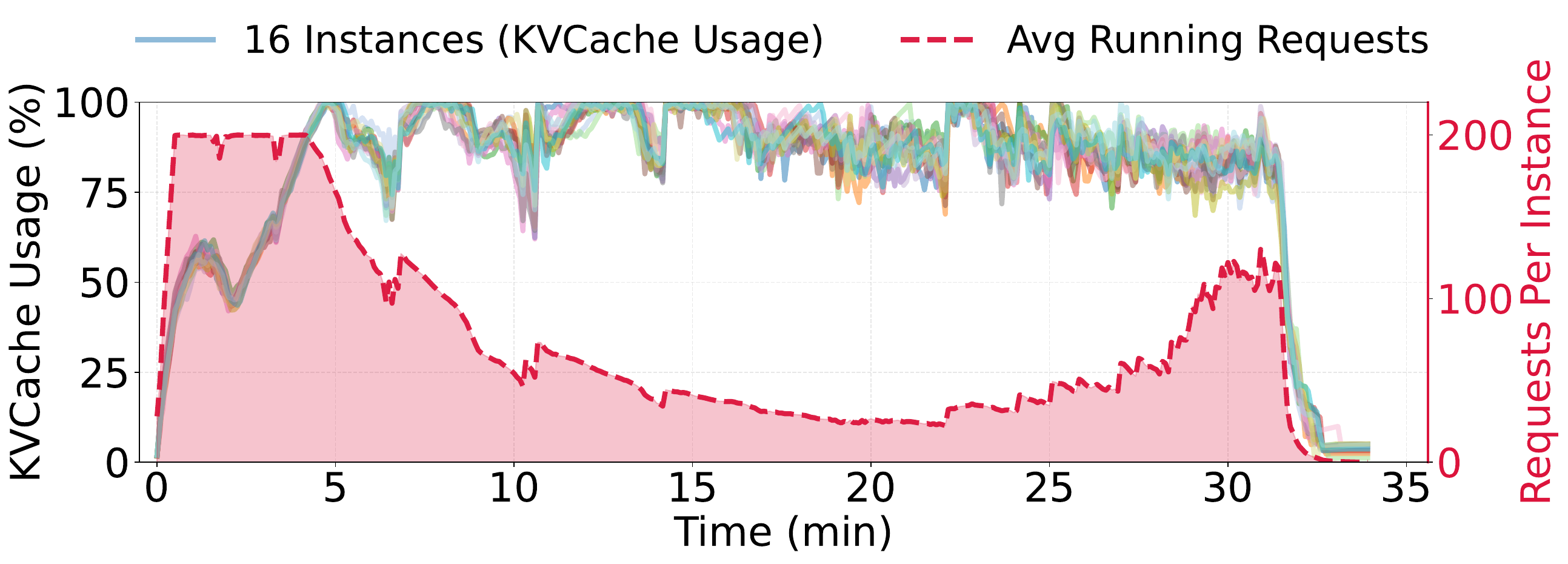}
\caption{KVCache utilization and average running requests with \systemname during a rollout iteration of the Qwen2-VL-72B task.}
\label{fig:seer_kv_util}
\vspace{-1em}
\end{figure}

To analyze \systemname's performance advantages, we examine tail latency in the rollout process. We define \textbf{tail requests} as the last 10\% of requests to complete in a synchronous system rollout, and \textbf{tail time} as the time spent \textbf{solely} processing these tail requests. Figure~\ref{fig:tail_latency} shows the tail time and total rollout completion time, averaged across all iterations. 
The results demonstrate severe tail latency for memory-constrained tasks like Moonlight and Qwen2-VL-72B, where the last 10\% of requests consume up to 50\% of total time.

The tail latency problem stems from two primary causes. First, request queuing and preemption lead to delayed scheduling of long-output requests. Second, monolithic request groups result in load imbalance across instances. We provide quantitative evidence using the Qwen2-VL-72B task on veRL shown in Figure~\ref{fig:rollout_analysis}. In this rollout iteration, a total of 13,686 preemption events occurred. The last 5\% of completed requests have an average length in the top 15th percentile, yet their average execution start time is at 42\% of the total time. The load imbalance across instances is even more pronounced: the completion time difference between the earliest and latest instances accounts for 70\% of the total time, with each instance idle for an average of 1,580 seconds, representing 37\% of the total time. These scheduling and load imbalances result in severe tail latency and throughput degradation.

By leveraging group-aware context learning and fine-grained request scheduling, \systemname significantly reduces tail latency by 72\% to 94\%, thereby substantially improving system throughput. Figure~\ref{fig:seer_kv_util} illustrates the impact of these techniques, showing that \systemname substantially reduces the tail phase duration compared to the baseline shown in Figure~\ref{fig:kv_util}.

\subsection{Improvement Breakdown}
\label{sec:ablation}

\begin{table}[t]
\centering
\caption{Performance improvement breakdown across three RL tasks. Context Sched. denotes context-aware scheduling (\S\ref{sec:context_aware_scheduling}), and Grouped SD denotes adaptive grouped speculative decoding (\S\ref{sec:speculative_decoding}).}
\label{tab:improvement_breakdown}
\small
\begin{tabular}{lccc}
\toprule
\textbf{Method} & \textbf{Moonlight} & \textbf{Qwen2-VL-72B} & \textbf{Kimi-K2} \\
\midrule
Baseline & 1.00 & 1.00 & 1.00 \\
+ Divided Rollout & 1.41$\times$ & 1.42$\times$ & 1.16$\times$ \\
+ Context Sched. & 1.47$\times$ & 1.56$\times$ & 1.27$\times$ \\
+ Grouped SD & 1.90$\times$ & 2.04$\times$ & 1.53$\times$ \\
\bottomrule
\end{tabular}
\end{table}

We present a systematic ablation study to quantify the contribution of each optimization component to \systemname's overall performance. Table~\ref{tab:improvement_breakdown} demonstrates the cumulative end-to-end speedup obtained by incrementally integrating each major optimization component, where each set of experiments is executed on the 5th rollout iteration.

First, \systemname's divided rollout mechanism enables dynamic, fine-grained load balancing across instances, mitigating tail latency caused by inter-instance load imbalance and reducing preemption overhead within instances due to varying KVCache memory consumption. This optimization yields significant improvements for memory-constrained tasks, achieving up to 42\% throughput improvement. Building upon divided rollout, context-aware scheduling leverages learned intra-group length distributions to guide scheduling, providing up to 14\% additional throughput improvement. To further address the computational underutilization inherent in tail requests, \systemname incorporates adaptive grouped speculative decoding, which contributes an additional 26--48\% performance improvement over the scheduling optimizations alone.

\subsection{Extended Studies}
\label{sec:case_study}

\subsubsection{Context-Aware Scheduling}
\label{sec:ablation_of_context}

\begin{figure}[t]
\centering
\begin{subfigure}[b]{0.49\columnwidth}
    \centering
    \includegraphics[width=\textwidth]{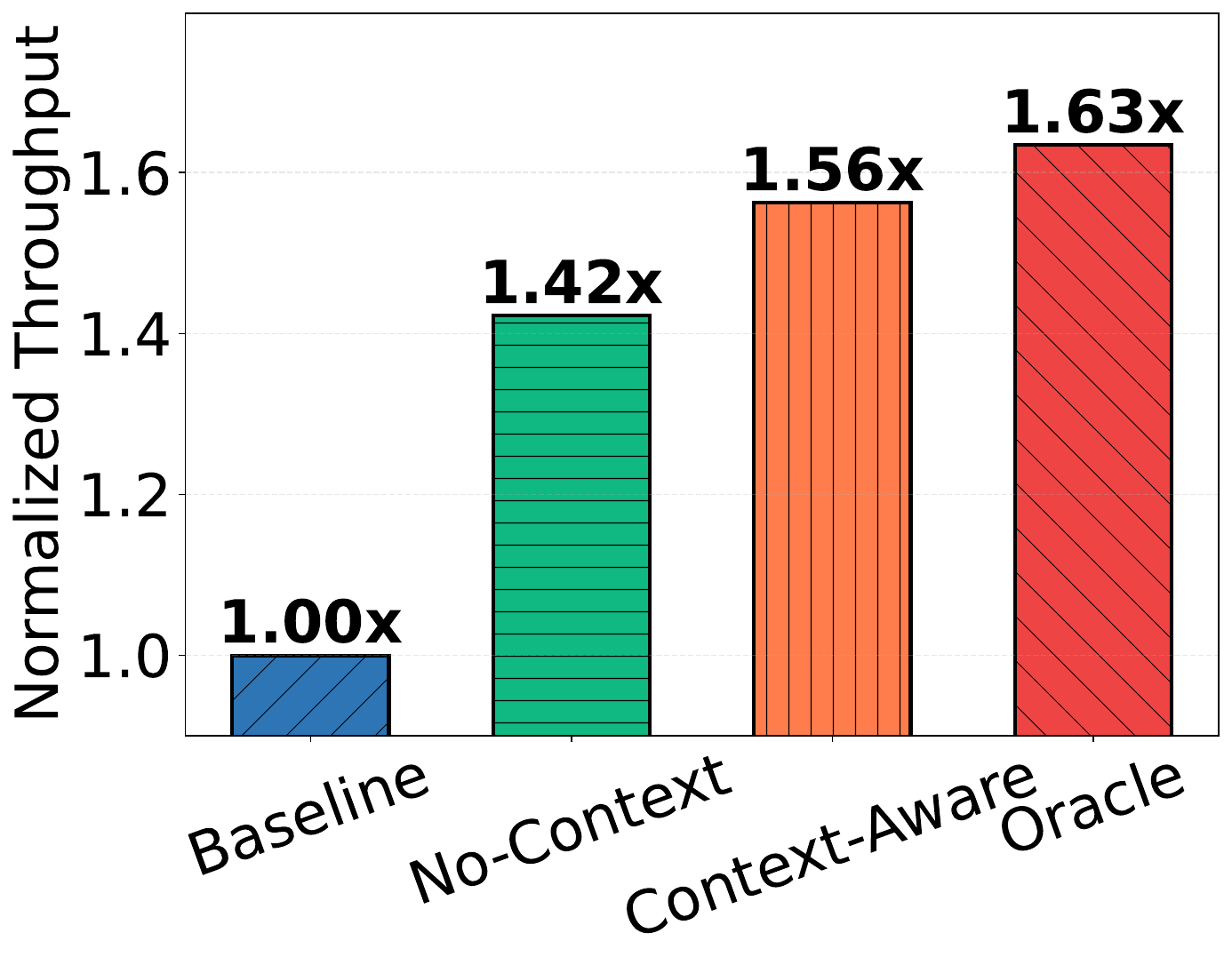}
    \caption{Normalized throughput.}
    \label{fig:length_ablation_throughput}
\end{subfigure}
\hfill
\begin{subfigure}[b]{0.49\columnwidth}
    \centering
    \includegraphics[width=\textwidth]{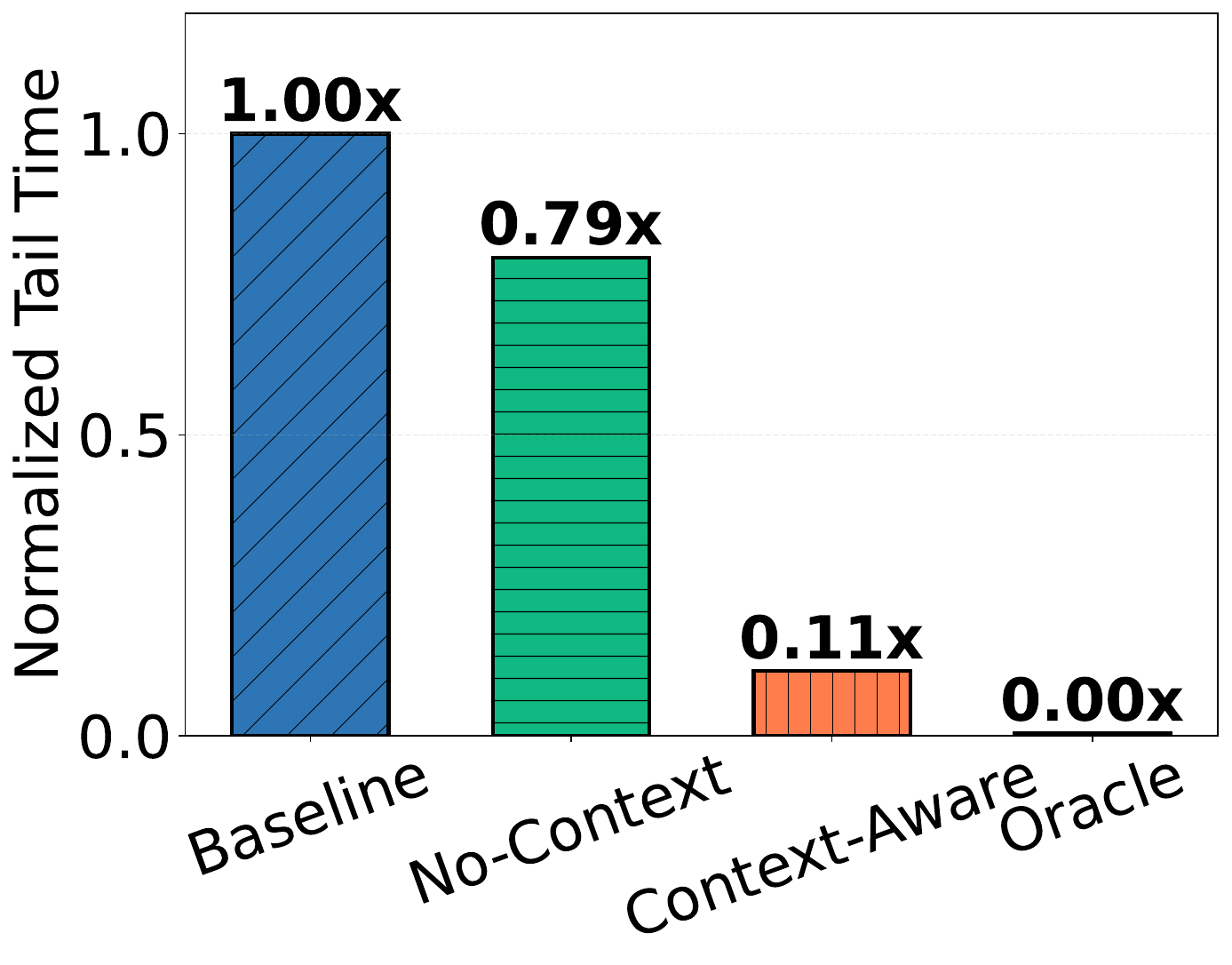}
    \caption{Normalized tail latency.}
    \label{fig:length_ablation_tail}
\end{subfigure}
\caption{Impact of length context on improving throughput and reducing tail latency (defined in \S\ref{sec:tail_latency}). No-Context applies only divided rollout without using length context to guide scheduling. Oracle obtains all output lengths in advance and applies the LFS strategy.}
\label{fig:length_ablation}
\vspace{-1em}
\end{figure}

To validate the effectiveness of length context, we conduct an ablation study by varying the length information available to \systemname's scheduler, as shown in Figure~\ref{fig:length_ablation}. Specifically, we compare: (1) No-Context, which applies only divided rollout without using length context to guide scheduling, and (2) Oracle, which obtains the actual output lengths of all requests in advance and replays the rollout iteration using these precise lengths to guide scheduling.

While divided rollout significantly improves throughput through dynamic load balancing, the tail latency problem persists, with tail latency reduced by only 21\% compared to the baseline. In contrast, context-aware scheduling leverages length prediction information from speculative requests to implement approximate longest-first scheduling, substantially reducing tail latency by 89\% compared to the baseline. Compared to Oracle, context-aware scheduling achieves 96\% of its throughput performance. This demonstrates that despite some performance degradation from prediction errors, context-aware scheduling still provides substantial benefits.

\subsubsection{Adaptive Grouped Speculative Decoding}
\label{sec:ablation_of_sd}

\begin{figure}[t]
\centering
\includegraphics[width=\columnwidth]{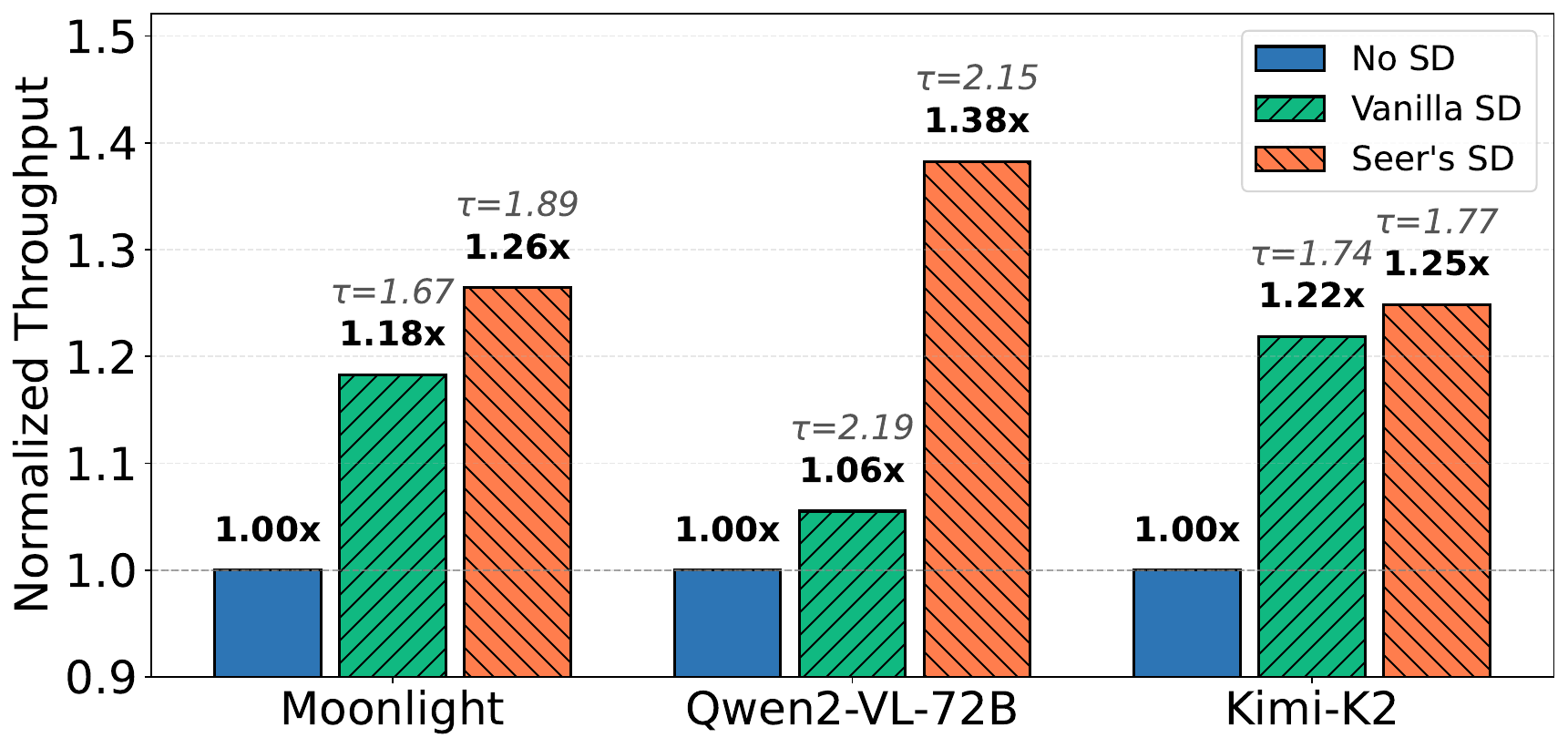}
\caption{Normalized throughput and mean acceptance length ($\tau$) of SD strategies across three tasks.}
\label{fig:sd_ablation}
\end{figure}

As discussed in \S\ref{sec:speculative_decoding}, the benefits of speculative decoding during rollout depend on multiple factors including batch size, draft model overhead, and draft accuracy. We conduct ablation experiments on different SD strategies based on a single rollout iteration executed on veRL. As shown in Figure~\ref{fig:sd_ablation}, \systemname's adaptive grouped SD consistently outperforms vanilla SD in throughput across all tasks, achieving up to 1.3$\times$ speedup. Furthermore, by leveraging group context, our approach improves the mean acceptance length by 0.22 compared to the CST-based SD method~\cite{oliaro2025suffixdecoding}, and exceeds MTP's performance. While vanilla SD with a small draft model achieves slightly higher mean acceptance length than our adaptive grouped SD, its excessive draft model overhead results in the lowest overall throughput gain.

\subsubsection{Non-Strictly Synchronous RL}
\label{sec:async_rl}

\begin{figure}[t]
\centering
\begin{subfigure}[b]{0.49\columnwidth}
    \centering
    \includegraphics[width=\textwidth]{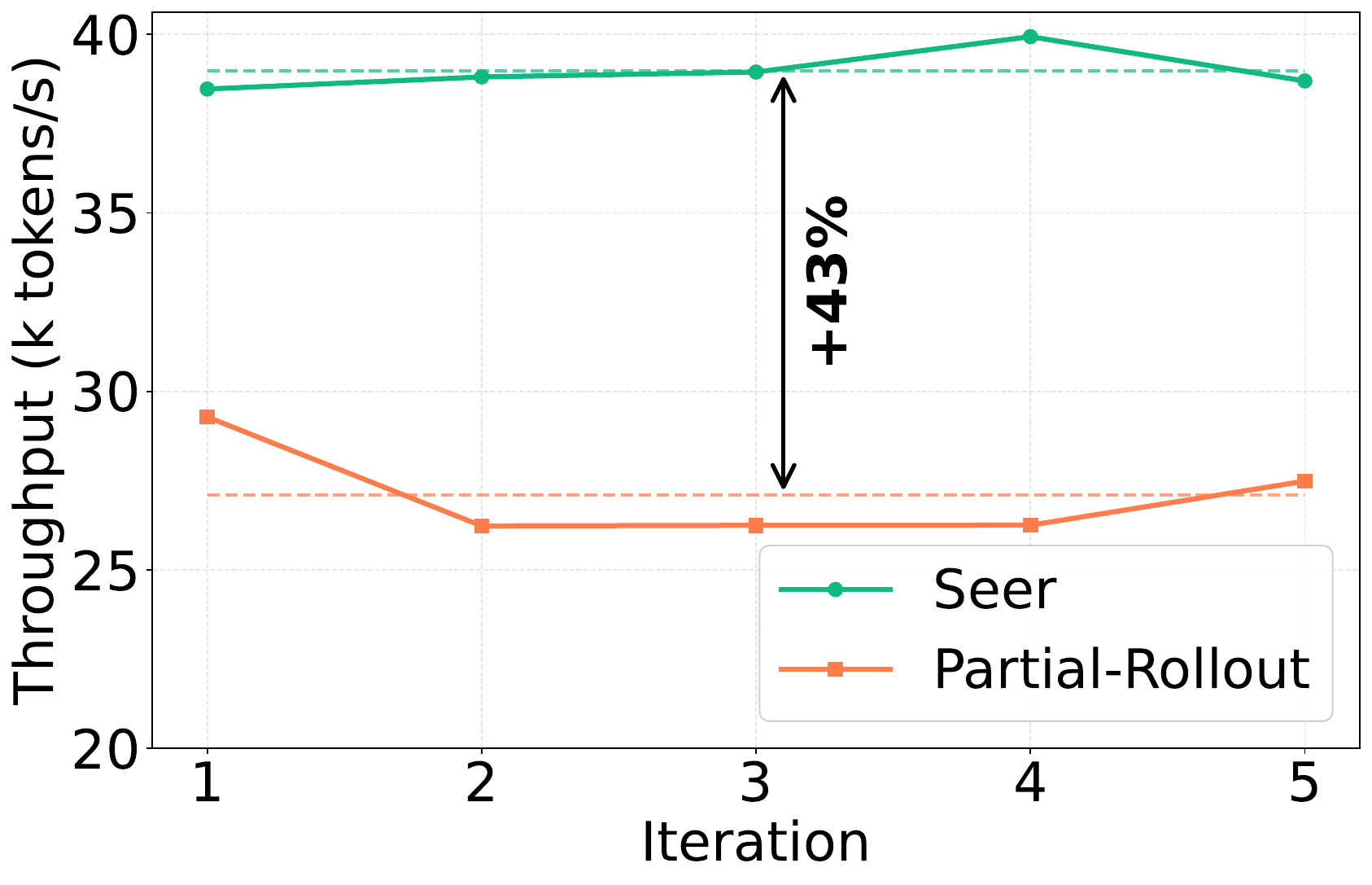}
    \caption{Rollout throughput.}
    \label{fig:exp_partial_e2e}
\end{subfigure}
\hfill
\begin{subfigure}[b]{0.49\columnwidth}
    \centering
    \includegraphics[width=\textwidth]{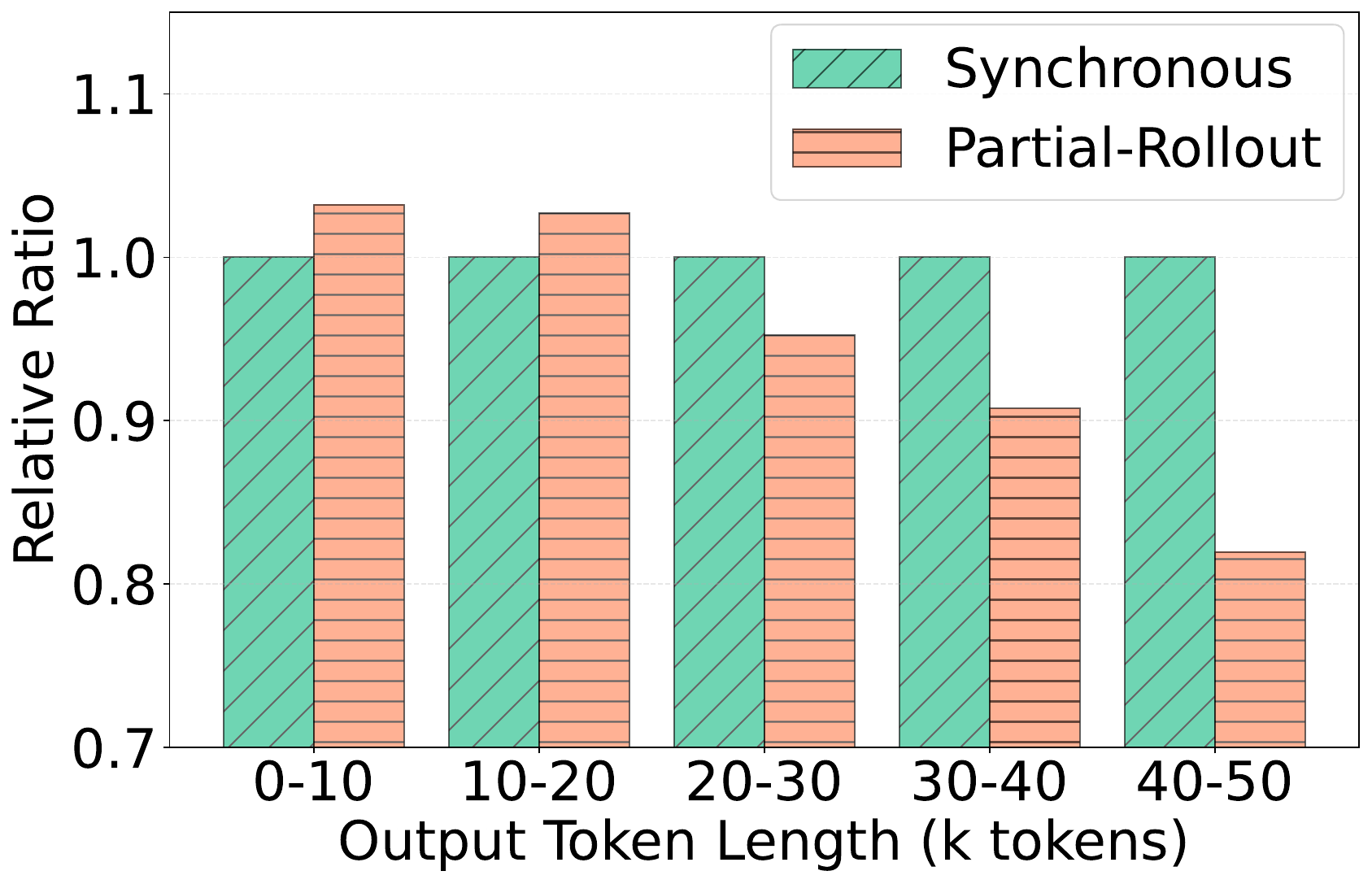}
    \caption{Output length distribution.}
    \label{fig:exp_partial_count}
\end{subfigure}
\caption{Rollout experiments comparing \systemname and Partial Rollout on the Qwen2-VL-72B workload.}
\label{fig:exp_partial}\
\vspace{-1.5em}
\end{figure}

Some RL training systems~\cite{zhou2025april,team2025kimik15,gao2025rollpacker} employ non-strictly synchronous RL, which defers or reschedules long-tail requests to subsequent iterations. While non-strictly synchronous RL can improve resource utilization, it introduces some degree of inconsistency with synchronous training algorithms.
Partial Rollout~\cite{zhou2025april,team2025kimik15} is a popular non-strictly synchronous method that over-issues requests during rollout and terminates the rollout phase once a predetermined number of requests are completed, with remaining requests executed with high priority in the next rollout iteration. In this section, we compare \systemname against Partial Rollout using the Qwen2-VL-72B workload. Following the experimental setup from APRIL~\cite{zhou2025april}, which applies Partial Rollout, we over-issue twice the number of requests.

As shown in Figure~\ref{fig:exp_partial_e2e}, \systemname achieves 43\% higher average throughput than Partial Rollout. This advantage stems from two factors. First, in memory-constrained long-CoT scenarios, Partial Rollout's doubling of request volume exacerbates preemption issues, preventing it from fully leveraging its higher degree of parallelism. Second, \systemname's adaptive grouped speculative decoding further enhances throughput for long-generation requests. Additionally, Figure~\ref{fig:exp_partial_count} shows the output length distributions of synchronous RL methods and Partial Rollout. Compared to the synchronous method, Partial Rollout generates a significantly lower proportion of requests with longer output lengths. This bias may negatively impact RL training and degrade model performance.
\section{Related Work}

\noindent\textbf{RL Frameworks for LLM Post-Training.} Numerous open-source RL frameworks~\cite{vonwerra2022trl,yao2023deepspeed,shen2024nemo,hu2024openrlhf,sheng2025hybridflow,wang2025reinforcement,slime_github} have been proposed to achieve both efficiency and usability, providing robust infrastructure support for both research and production RL workloads. OpenRLHF~\cite{hu2024openrlhf} and veRL~\cite{sheng2025hybridflow} seamlessly integrate training and inference engines, supporting various parallelism strategies to improve RL training efficiency. ROLL~\cite{wang2025reinforcement} targets large-scale RL training optimization, enabling flexible resource allocation and heterogeneous task scheduling. Slime~\cite{slime_github} balances high performance with flexibility, supporting highly customizable data generation pipelines. These works primarily focus on overall RL training workflow orchestration, while the long-tail problem in the rollout phase remains unaddressed.

\noindent\textbf{On-Policy RL Optimization.} Some efforts have been made to optimize the RL workflow at both the system and algorithmic levels. RealHF~\cite{mei2024real} dynamically reallocates model parameters and memory budgets to optimize utilization. RLHFuse~\cite{zhong2025optimizing} proposes stage fusion to overlap reward and experience computation with the rollout long tail, thereby improving resource utilization. RLBoost~\cite{wu2025rlboost} utilizes preemptible fragmented GPU resources to accelerate rollout at low cost. Nevertheless, these approaches do not adequately address the long-tail latency problem. For the long-tail problem, speculative decoding~\cite{leviathan2023fast,li2024eagle1,li2024eagle2,li2025eagle3,cai2024medusa,liu2024deepseekv3,fu2024break,oliaro2025suffixdecoding,hu2025sam} is regarded as an effective optimization technique. However, these methods suffer from either high draft overhead or low accuracy in RL rollout scenarios due to dynamically changing batch sizes and continuously evolving target LLM weights. RhymeRL~\cite{he2025history} and SPEC-RL~\cite{liu2025spec} propose utilizing historical sequences generated from previous RL iterations as references to accelerate decoding. However, in most state-of-the-art model rollout scenarios, different iterations sample different data to improve generalization, leaving no historical sequences to reuse. In contrast, \systemname exploits the similarity among requests within the same group, eliminating the need to rely on repeated sequences across iterations. Meanwhile, \systemname employs an adaptive speculation policy based on a disaggregated architecture, achieving enhanced accuracy while maintaining low draft overhead.

\noindent\textbf{Off-Policy RL Optimization.} Recently, many works~\cite{zhong2025streamrl,fu2025areal,han2025asyncflow,sheng2025laminar,he2025history,zhou2025april,gao2025rollpacker,piche2025pipelinerl,lu2025part} have proposed RL workflows that do not maintain complete logical synchronization, trading a degree of off-policy behavior for improved RL efficiency. StreamRL~\cite{zhong2025streamrl}, Areal~\cite{fu2025areal}, RhymeRL~\cite{he2025history}, and Laminar~\cite{sheng2025laminar} propose asynchronous training approaches to improve resource utilization. In asynchronous RL, training and rollout are completely decoupled: rollout workers continuously generate new outputs without waiting, while training workers update the model whenever a batch of samples is collected. Other works relax the synchronization constraints of synchronous RL to achieve speedup, referred to as non-strictly synchronous RL. Partial Rollout~\cite{team2025kimik15,zhou2025april} defers long-tail requests to continue execution in the next rollout iteration. RollPacker~\cite{gao2025rollpacker} proposes tail batching, which schedules long-tail requests into a few designated long iterations, effectively reducing GPU idle time. While these techniques improve RL efficiency, the off-policyness inherent in these algorithms introduces risks of instability and accuracy degradation, limiting their applicability in research and production environments where algorithmic fidelity is critical. \systemname significantly reduces rollout tail latency through fine-grained scheduling of group-aware context learning within the rollout phase, while maintaining consistency with on-policy algorithms.
\section{Conclusion}

In this paper, we present \systemname, a synchronous RL system that accelerates the rollout process through group-aware context learning. With divided rollout, a fine-grained and dynamically load-balanced scheduling approach, \systemname leverages the similarity among requests within the same group in GRPO-like algorithms to achieve efficient scheduling and speculative decoding, while strictly maintaining consistency with on-policy RL algorithms. Experimental results demonstrate that \systemname achieves up to 2.04$\times$ throughput improvement and up to 94\% reduction in long-tail latency compared to the current state-of-the-art RL framework.

\bibliographystyle{plain}
\bibliography{main}

\cleardoublepage
\appendix
\section{Implementation Details of \systemname}

\subsection{Context-Aware Scheduling Algorithm}

Algorithm~\ref{alg:context_aware_scheduling} presents the context-aware scheduling workflow built on top of divided rollout. The algorithm is invoked continuously by \systemname's global scheduler, each time returning a scheduling decision $(r^\star, i^\star)$ that assigns the selected request $r^\star$ to an inference instance $i^\star$, until all requests are completed.

\begin{algorithm}[h]
\small
\caption{Context-Aware Scheduling based on Divided Rollout}\label{alg:context_aware_scheduling}
\begin{algorithmic}[1]
\Require {
    Active requests $\mathcal{R}=\{r_{g,i}\}$ grouped by prompt $g$;\;
    group-level length estimates $\{\widehat{L}_g\}$;\;
    inference instances $\mathcal{I}$ with KV-usage telemetry.
}
\Ensure A scheduling decision $(r^\star, i^\star)$ with $r^\star \in \mathcal{R}$ and $i^\star \in \mathcal{I}$.

\ForAll{$r_{g,i}\in\mathcal{R}$}
    \If{$r_{g,i}$ is finished}
        \State $\widehat{L}_g\leftarrow\Call{UpdateEstimate}{\widehat{L}_g, L_{g,i}}$
        \State remove $r_{g,i}$ from $\mathcal{R}$
    \ElsIf{$r_{g,i}$ is the group's speculative request}
        \State keep in high-priority queue $\mathsf{Q}_{\text{spec}}$
    \Else
        \State add to low-priority candidate set $\mathsf{C}_{\text{rest}}$
    \EndIf
\EndFor

\State $r^\star \gets \textbf{None}$
\If{$\neg \Call{IsEmpty}{\mathsf{Q}_{\text{spec}}}$}
    \State $r^\star \gets \Call{PickSFS}{\mathsf{Q}_{\text{spec}}}$ \Comment{smallest generated length first}
\ElsIf{$\neg \Call{IsEmpty}{\mathsf{C}_{\text{rest}}}$}
    \State $r^\star \gets \Call{PickLFS}{\mathsf{C}_{\text{rest}}}$ \Comment{largest $\widehat{L}_g$ first}
\Else
    \State \Return \texttt{all requests are finished}
\EndIf

\State $
\begin{aligned}
r^\star.\mathit{max\_tokens} \gets {} &
\min\!\big(chunk\_size, \\
& \, r^\star.\mathit{ori\_max\_tokens}-r^\star.\mathit{generated\_tokens}\big)
\end{aligned}
$

\State $i^\star \gets \Call{SelectInstance}{\mathcal{I}, r^\star.max\_chunk\_tokens, \text{KV-usage}}$
\If{$i^\star \neq \textbf{None}$}
    \State \Return $(r^\star, i^\star)$
\EndIf

\State \Return \texttt{no available instance for this cycle}

\end{algorithmic}
\end{algorithm}

\subsection{Workflow of Distributed Grouped Draft Server}

\begin{figure}
\begin{center}
    \includegraphics[width=\linewidth]{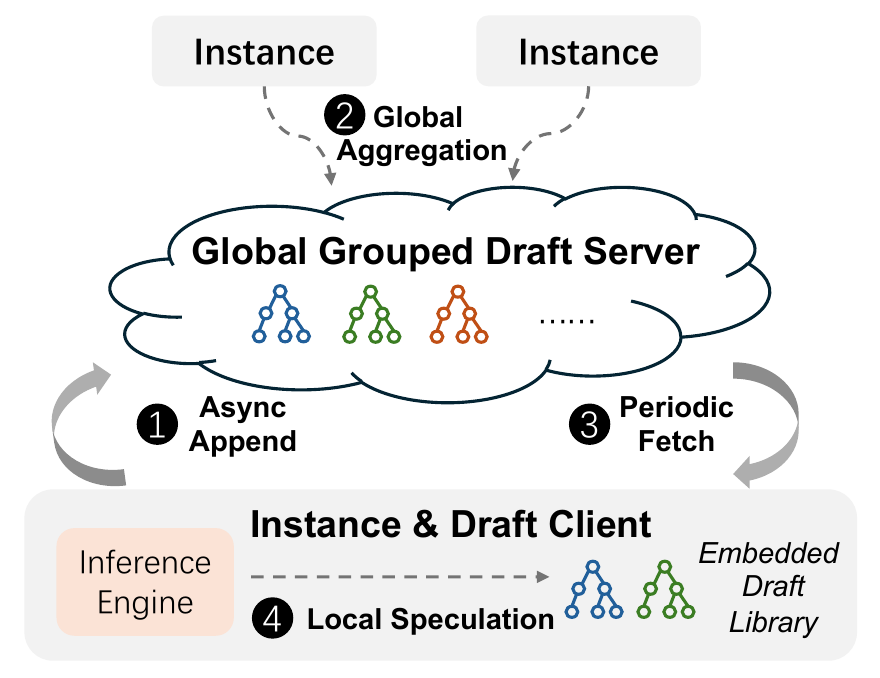}
    \caption{The distributed grouped draft server.}
    \label{fig:draft_server_appendix}
\end{center}
\end{figure}

To minimize speculative decoding latency in the critical path, Distributed Grouped Draft Server (DGDS) adopts asynchronous updates and employs a distributed master-worker architecture for global sharing of grouped context, as illustrated in Figure~\ref{fig:draft_server_appendix}. The system operates through four key steps, with core APIs listed in Table~\ref{tab:draft_server_api} and Table~\ref{tab:draft_client_api}:

\noindent\textit{\underline{1) Asynchronous Append:}} Each inference instance runs an independent process to handle output tokens. When newly generated tokens are produced, the instance invokes the \texttt{update\_cst} API to send them to DGDS, identified by \texttt{group\_id}. To reduce communication overhead, each request batches a fixed number of tokens before sending updates, which has negligible impact on draft token quality.

\noindent\textit{\underline{2) Global Aggregation:}} DGDS aggregates token updates from requests belonging to the same group. To prevent cross-request interference, DGDS isolates updates by \texttt{request\_id}, mapping each new token only to the corresponding local path in the CST.

\noindent\textit{\underline{3) Periodic Fetch:}} Each inference instance embeds a draft client as a library component that periodically synchronizes the latest CST from DGDS. The client registers active request groups via \texttt{register\_group} and then periodically invokes \texttt{fetch\_cst} to retrieve the latest CSTs for these groups. To reduce communication overhead, the client supports incremental synchronization based on local cache states.

\noindent\textit{\underline{4) Local Speculation:}} Inference instances perform speculation based on their local CSTs by invoking the \texttt{batch\_speculate} API. The local CSTs aggregate paths from all requests within the same group, enabling instances to share contextual statistics and obtain higher-quality draft tokens.

\begin{table*}[t]
\centering
\caption{Grouped draft server API.}
\label{tab:draft_server_api}
\small
\begin{tabular}{p{3.2cm}p{5.5cm}p{7.5cm}}
\hline
\textbf{Operation} & \textbf{Description} & \textbf{Parameters} \\
\hline
\texttt{update\_cst} & Append generated tokens from a specific request to the compressed suffix tree & \texttt{const string\& group\_id, int request\_id, int prev\_token\_count, const vector<int>\& new\_tokens} \\
\texttt{fetch\_cst} & Fetch incremental draft contexts of request groups based on their current cache states (if have) & \texttt{const vector<string>\& group\_ids, const vector<DraftCacheInfo>\& draft\_cache\_infos} \\
\hline
\end{tabular}
\end{table*}

\begin{table*}[t]
\centering
\caption{Draft client API.}
\label{tab:draft_client_api}
\small
\begin{tabular}{p{3.2cm}p{5.5cm}p{7.5cm}}
\hline
\textbf{Operation} & \textbf{Description} & \textbf{Parameters} \\
\hline
\texttt{register\_group} & Register a new request group for draft fetching with TTL & \texttt{const string\& group\_id, int ttl\_seconds} \\
\texttt{batch\_speculate} & Generate speculative tokens for multiple requests via zero-copy memory access: reads input token patterns and writes predicted tokens directly to inference engine memory buffers & \texttt{const vector<string>\& group\_ids, const vector<size\_t>\& buffer\_offsets, void* pattern\_buffer\_ptr, void* output\_buffer\_ptr, const vector<SpeculationArgs>\& speculation\_args} \\
\hline
\multicolumn{3}{l}{\textbf{SpeculationArgs}: \texttt{max\_spec\_tokens} (int), \texttt{pattern\_lookup\_max} (int), \texttt{pattern\_lookup\_min} (int), \texttt{top\_k} (int)} \\
\hline
\end{tabular}
\end{table*}
\cleardoublepage

%%%%%%%%%%%%%%%%%%%%%%%%%%%%%%%%%%%%%%%%%%%%%%%%%%%%%%%%%%%%%%%%%%%%%%%%%%%%%%%%
\end{document}